\documentclass[preprint,notoc]{JHEP3} % 10pt is ignored!
\usepackage{epsfig,multicol}
\usepackage{amssymb}
\usepackage{graphics}
\usepackage{cite}

\usepackage{amsmath,latexsym}
\newcommand{\Ghbs}{\ensuremath{\Gamma(h\to q\,{q'})}}
\newcommand{\Bhbs}{\ensuremath{B(h\to q\,{q'})}}
\newcommand{\Bhbsmax}{\ensuremath{B^{\rm max}(h\to q\,{q'})}}
\newcommand{\Bhzbs}{\ensuremath{B(h^0\to q\,{q'})}}
\newcommand{\Ghzbs}{\ensuremath{\Gamma(h^0\to q\,{q'})}}
\newcommand{\Ghzbsmax}{\ensuremath{\Gamma^{\rm max}(h^0\to q\,{q'})}}
\newcommand{\Bhzbsmax}{\ensuremath{B^{\rm max}(h^0\to q\,{q'})}}
\newcommand{\BHAbsmax}{\ensuremath{B^{\rm max}(H^0/A^0\to q\,{q'})}}
\newcommand{\bsg}{\ensuremath{b\to s\gamma}}
\newcommand{\Bbsg}{\ensuremath{B(b\to s\gamma)}}
\newcommand{\ma}{\ensuremath{m_{A^0}}}
\newcommand{\tb}{\ensuremath{\tan\beta}}
\newcommand{\GeV}{\mbox{ GeV}}
\newcommand{\TeV}{\mbox{ TeV}}

\newcommand{\sbottom}{\ensuremath{\tilde{b}}}
\newcommand{\sstrange}{\ensuremath{\tilde{s}}}
\newcommand{\squark}{\ensuremath{\tilde{q}}}
\newcommand{\stopp}{\ensuremath{\tilde{t}}}

\newcommand{\gsim}{\mbox{ \raisebox{-4pt}{${\stackrel{\textstyle >}{\sim}}$} }}

\newcommand{\mb}{m_b}
\newcommand{\mt}{m_t}
\newcommand{\mw}{M_W}
\newcommand{\mHp}{m_{H^\pm}}
\newcommand{\mA}{m_{A^0}}
\newcommand{\mg}{m_{\tilde{g}}}

\newcommand{\PDG}{Hagiwara:2002fs}
\newcommand{\SUSY}{Nilles:1984ex,Haber:1985rc,Lahanas:1987uc,Ferrara87}
\newcommand{\TESLA}{Aguilar-Saavedra:2001rg}
\newcommand{\LHC}{Atlas,CMS}
\newcommand{\GuaschSola}{Guasch:1999jp,Bejar:2001sj}
\newcommand{\Dabels}{Yamada:1994kj,Chankowski:1994er,Dabelstein:1995hb,Dabelstein:1995js}
\newcommand{\BurasGabbiani}{Gabbiani:1996hi,Misiak:1997ei}
\newcommand{\bsgold}{Barbieri:1993av,Garisto:1993jc,Diaz:1994fc,Borzumati:1994zg,Bertolini:1995cv}
\newcommand{\Santi}{Bejar:2000ub,Bejar:2001sj,Bejar:2003em}
\newcommand{\bsgexp}{Alam:1995aw,Barate:1998vz,Ahmed:1999fh,Abe:2001hk,Chen:2001fj,Aubert:2002pd}
\newcommand{\bsgthdm}{Ciuchini:1998xe,Borzumati:1998tg,Borzumati:1998nx}
\newcommand{\hbbQCD}{Braaten:1980yq,Sakai:1980fa,Inami:1981qp,Drees:1990du,Drees:1990dq}
\newcommand{\FeynArts}{Kublbeck:1990xc,Hahn:1998yk,FAFCuser}

\title{Higgs Boson Flavor-Changing Neutral Decays into Bottom Quarks
  in Supersymmetry}
\author{
Santi B\'ejar $^{a}$, Francesc Dilm{\'e} $^{b}$, Jaume Guasch $^{c}$, Joan Sol\`a $^{b,d}$\\
$^{a}\,$ Grup de F{\'\i}sica Te{\`o}rica and IFAE, Universitat
Aut{\`o}noma de Barcelona, \\
\mbox{\hspace{+0.2cm}}  E-08193, Bellaterra,
    Barcelona, Catalonia, Spain\\
$^{b}\,$ Dep. Estructura i Constituents de la Mat\`eria,
Universitat de Barcelona, \\ \mbox{\hspace{+0.2cm}} Diagonal 647,
E-08028, Barcelona, Catalonia, Spain\\
$^{c}\,$ Theory Group LTP, Paul Scherrer Institut, CH-5232
Villigen PSI, Switzerland\\
 $^{d}$ C.E.R. for
Astrophysics, Particle Physics and Cosmology
\thanks{Associated with Instituto de Ciencias del
Espacio-CSIC.}}

\preprint{PSI-PR-03-18\\ UB-ECM-PF 04/02\\ hep-ph/0402188}

\abstract{We analyze the maximum branching ratios for the Flavor
Changing Neutral Current (FCNC) decays of the neutral Higgs bosons
of the Minimal Supersymmetric Standard Model (MSSM) into bottom
quarks, $h\to b\bar{s}$ ($h=h^0,H^0,A^0$). We consistently
correlate these decays with the radiative B-meson decays $(b\to
s\gamma)$. A full-fledged combined numerical analysis is performed
of these high-energy and low-energy FCNC decay modes in the MSSM
parameter space. Our calculation shows that the available data on
$\Bbsg$ severely restricts the allowed values of $B(h\rightarrow
{b}\,\bar{s})$.  While the latter could reach a few percent level
in fine-tuned scenarios, the requirement of naturalness reduces
these FCNC rates into the {modest range $B(h\rightarrow
{b}\,\bar{s})\sim 10^{-4}-10^{-3}$}. We find that the bulk of the
MSSM contribution to $B(h\rightarrow {b}\,\bar{s})$ could
originate from the {strong supersymmetric sector}. The
maximum value of {the} FCNC rates {obtained} in
this paper disagree significantly with {recent}
(over-)estimates existing in the literature. Our results are
still encouraging because they show that the FCNC modes $h\to
b\bar{s}$ can be competitive with other Higgs boson signatures
and could play a helpful complementary role to identify the
supersymmetric Higgs bosons, particularly the lightest CP-even
state in the critical LHC mass region $m_{h^0}\simeq 90-130\GeV$.}


\begin{document}

%%%%%%%%%%%%%%%%%%%%%%%%%%%%%%%%%%%%%%%%%%%%%%%%%%%%%%%%%%%
\section{Introduction}
\label{sect:introduction}

Experimentally, processes involving Flavor Changing Neutral
Current (FCNC) have been shown to have extremely low
rates~\cite{\PDG}. Theoretically, their rareness can be explained
by the GIM mechanism~\cite{Glashow:1970gm}, which is related to
the unitarity of the mixing matrices between quarks. The Minimal
Standard Model (SM) embeds naturally the GIM mechanism, due to
the presence of only one Higgs doublet giving mass simultaneously
to the down-type and the up-type quarks, and as a result no
tree-level FCNCs interactions appear. FCNCs are radiatively
induced, and are therefore automatically small. The addition of
further Higgs doublets to the SM in the most general way
introduces tree-level FCNC interactions, which would predict
significant FCNC rates. {However}, by introducing an
\textit{ad-hoc} discrete symmetry these interactions are
forbidden. This gives rise to two classes of Two-Higgs-Doublet
Models (2HDM) which avoid FCNCs at the tree-level, known
conventionally as type~I and type~II 2HDMs~\cite{Hunter}.

Supersymmetry (SUSY)~\cite{\SUSY}, on the other hand, provides an
appealing extension {of} the SM, which unifies the
fermionic and bosonic degrees of freedom of the fundamental
particles, and provides a natural solution to the hierarchy
problem. The search for SUSY particles has been one of the main
programs of the past experiments in high energy physics (LEP,
SLD, Tevatron), and continues {to play a central role} in
the present accelerator experiments (Tevatron II), and in the
planning of future experimental facilities, {{like the
LHC and the LC}~\cite{\LHC,\TESLA}. The Minimal Supersymmetric
Standard Model (MSSM) is the simplest extension of the SM which
includes SUSY, {and for this reason its testing will be
one of the most prominent aims of these powerful experiments.}

Complementary to direct searches, one can also look for effects
of particles beyond the SM by studying their radiative effects.
{Much work along these lines has been made over the past two
decades. FCNCs may play an important role here because they are
essentially loop-induced. Hence SM and non-SM loops enter the FCNC
observables at the same order of perturbation theory, and new
physics competes on the same footing with SM physics to generate a
non-vanishing value for these rare processes. It may well be that
the non-SM effects are dominant and become manifest. Conversely,
it may happen that they become highly constrained. The power of
FCNC observables can be gauged}
{e.g.} by the implications of the bottom-quark rare decay
$\bsg$: the experimentally measured allowed range
$\Bbsg=(3.3\pm0.4)\times 10^{-4}$~\cite{\bsgexp,Hagiwara:2002fs}
{may impose tight constraints on extensions of the SM. For
example, it implies a lower bound on the charged Higgs boson mass
$\mHp\gsim 350\GeV$ in general type II
2HDMs~\cite{\bsgthdm,Gambino:2001ew}.}

The most general MSSM includes tree-level FCNCs among the extra
predicted particles, which induce one-loop FCNC interactions among
the SM particles. Given the observed smallness of these
interactions, tree-level SUSY FCNCs are usually avoided by
including one of the two following assumptions: either the SUSY
particle masses are very large, and their radiative effects are
suppressed by the large SUSY mass scale; or the soft SUSY-breaking
squark mass matrices are aligned with the SM quark mass matrix, so
that both mass matrices are simultaneously diagonal. However, if
one looks closely, one soon realizes that the MSSM does not only
include the possibility of tree-level FCNCs, but it actually
\textit{requires} their existence~\cite{Duncan:1983iq}. Indeed,
the requirement of $SU(2)_L$ gauge invariance means that the
up-left-squark mass matrix can not be simultaneously diagonal to
the down-left-squark mass matrix, and therefore these two matrices
can not be simultaneously diagonal with the up-quark and the
down-quark mass matrices, that is, unless both of them are
proportional to the identity matrix. {But even then we
could not take such possibility too seriously, for the radiative
corrections would produce non-zero elements in the non-diagonal
part of the mass matrix. All in all, we naturally expect
tree-level FCNC interactions mediated by the SUSY partners of the
SM particles. The potentially largest FCNC interactions are those
originating from the strong supersymmetric (SUSY-QCD) sector of
the model (viz. those interactions involving the
squark-quark-gluino couplings), and in this paper we mainly
concentrate on them. These couplings induce FCNC loop effects on
more conventional fermion-fermion interactions, like e.g. the
gauge boson-quark vertices $Vqq'$.}

{Of course, low energy meson physics puts tight
constraints on the possible value of the FCNC couplings,
especially for the first and second generation squarks which are
sensitive to the data on} $K^0-\bar{K}^0$ and
$D^0-\bar{D}^0$~\cite{\BurasGabbiani}. The third generation system is,
in principle, much loosely constrained, since present data on
$B^0-\bar{B}^0$ mixing still leaves a large room for FCNCs, and
the most stringent constraints are given by the $\Bbsg$
measurement~\cite{\bsgexp}. Therefore the relevant FCNC gluino
coupling $\delta_{23}$ \cite{\BurasGabbiani} (see {Section
\ref{sect:numerical}}) is not severely bound at present. The lack
of tight FCNC constraints in the top-bottom quark doublet enables
the aforementioned lower bound on the charged Higgs boson mass in
the MSSM to be easily avoided, to wit: by arranging that the
SUSY-electroweak ({{SUSY-EW}) contribution to $\Bbsg$ from
the top-squark/chargino loops screens partially the charged Higgs
boson contribution. This situation can be naturally fulfilled if
the higgsino mass parameter ($\mu$) and the soft SUSY-breaking
top-squark trilinear coupling ($A_t$) satisfy the relation
$\mu\,A_t < 0$~\cite{\bsgold}.

The FCNC gluino interactions also induce large contributions to
$\Bbsg$. {It should however be noted} that the leading
contributions to the {$Vqq'$ FCNC interactions from the third
quark generation} correspond to a \textit{double insertion} term,
in which the squarks propagating in the loop suffer a double
mutation: a flavor conversion and a chirality transition. This
fact has been demonstrated in the $\Bbsg$ observable
itself~\cite{Borzumati:1999qt}, as well as {in} the FCNC rare
decay width $\Gamma(t\to cg)$~\cite{\GuaschSola}. As a
consequence, the loose limits on the third generation FCNC
interactions derived under the assumption that the leading terms
contributing to $\bsg$ correspond to the \textit{single particle
insertion approximation}~\cite{\BurasGabbiani} are not valid, and
more complex expressions must be taken into
account~\cite{Besmer:2001cj}.

Concerning the FCNC interactions of Higgs bosons with third
generation quarks, it was demonstrated long ago~\cite{\GuaschSola}
that the leading term corresponds to a \textit{single particle
insertion approximation}, which produces a flavor {change} in the
internal squark loop propagator, since in this case the chirality
change can already take place at the squark-squark-Higgs boson
interaction vertex. Adding this to the fact that the Higgs bosons
({in contrast to gauge bosons}) have a privileged coupling
to third generation quarks, one might expect that the FCNC
interactions of the type quark-quark-Higgs bosons in the MSSM
{become highly strengthened with respect to the SM prediction}.
This was already proven in the rare decay channels $\Gamma(t\to
ch)$~\cite{\GuaschSola} ($h$ being any of the neutral Higgs bosons
of the MSSM $h\equiv h^0,H^0,A^0$), where the maximum rate of the
SUSY-QCD induced branching ratio was found to be $BR(t\to
ch)\simeq 10^{-5}$, eight orders of magnitude above the SM
expectations $BR(t\to cH^{SM})\simeq 10^{-13}$. Similar
enhancement factors have been found in the top-quark-Higgs boson
interactions in other extensions of the SM~\cite{\Santi}.

{From the experience of the previous calculations
with the top quark, we expect similar enhancements in the FCNC
interactions of the MSSM Higgs bosons with the bottom quark.
Indeed, the purpose of this paper is to quantify, in a reliable
way, the MSSM expectations on the FCNC Higgs boson decay modes}
\begin{equation}\label{hFCNC}
h\rightarrow {b}\,\bar{s}\,, \ \ \ h\rightarrow \bar{b}\,s\ \ \ \
\ \ (h=h^0,H^0,A^0)\,.
\end{equation}
There are other FCNC decay modes involving light quarks. However,
only these bottom quark channels are relevant, as the remaining
FCNC decays into light quarks are negligible in the MSSM.
Moreover, the FCNC decays of Higgs bosons into bottom quarks are
specially interesting as they can provide an invaluable tool to
discriminate among different extended Higgs boson scenarios in
the difficult LHC range $90<m_h<130\GeV$~\cite{\LHC}.

{In this paper we present what we believe is the first realistic
estimate of the SUSY-QCD  contributions to the FCNC branching
ratios of the MSSM Higgs bosons into bottom quark. Specifically,
we compute}
\begin{equation}
\Bhbs=\frac{\Gamma(h\to q\,q')}{\Gamma(h\to X)}\equiv
\frac{\Gamma(h\rightarrow {b}\,\bar{s})+\Gamma(h\rightarrow
\bar{b}\,s)}{\sum_i \Gamma(h\to X_i)} \label{eq:hbs-def}
\end{equation}
{for the three Higgs bosons of the MSSM, $h=h^0,H^0,A^0$, where
$\Gamma(h\to X)$ is the -- consistently computed -- total width
in each case. The maximization process of the above branching
ratios in the MSSM parameter space is performed on the basis of a
simultaneous analysis of the relevant partial decay widths and of
the branching ratio of the low-energy FCNC process $b\to
s\,\gamma$, whose value is severely restricted by
experiment~\cite{\bsgexp}. It turns out that the maximum FCNC
rates that we find disagree quite significantly with some
simplified estimates that have recently appeared in the
literature~\cite{Madrid}. According to these authors the FCNC
decay rate of some MSSM Higgs bosons into bottom quarks can reach
the level  $\Bhbs\sim 25\%$. We find this value untenable, even
more given the fact that in Ref.~\cite{Madrid} no attempt is made
to verify the restrictions of the parameter space imposed by the
low energy data on $\Bbsg$.\footnote{See also
Ref.~\cite{Demir:2003bv} for a
  combined analysis of flavor-violating and CP-violating MSSM couplings.}

{The structure of the paper is as follows.
In Section \ref{sect:partialwidths} we estimate the expected
branching ratios and describe the structure of
Eq.\,(\ref{eq:hbs-def}) in the MSSM in more detail; in Section
\ref{sect:numerical} we present the numerical analysis, and in
Section \ref{sect:conclusions} we deliver our conclusions.}

\section{Partial widths and branching ratios}
\label{sect:partialwidths}

{First of all let us estimate the branching ratio
(\ref{eq:hbs-def}) in the SM.  It is not necessary to perform a
detailed calculation to suspect that it is rather small. Using
dimensional analysis, power counting, CKM matrix elements and
dynamical features we expect that the maximum branching ratio of
the SM Higgs boson $H^{SM}$ into bottom quark is at most of
order\,\footnote{See Ref.\,\cite{Bejar:2003em} for details on
similar estimates, like that of $BR(H^{SM}\rightarrow
t\,\bar{c})$.}
\begin{equation}
    BR(H^{SM}\rightarrow b\,\bar{s})
    \sim\left(\frac{|V_{ts}|}{16\pi^2}\right)^2\,\alpha_{W}\
    G_F\,\left(\frac{m_H^4}{m_b^2}\right)\lesssim 10^{-7}\ \ \
    (\,{\rm if}\ m_H<2\,M_W)\,.
    \label{estimateBRSM}
\end{equation}
Here $G_F$ is Fermi's constant and $\alpha_{W}=g^2/4\pi$, $g$
being the $SU(2)_L$ weak gauge coupling. This result should hold
for a SM Higgs boson mass $m_H<2\,M_W$, and in particular in the
critical LHC region $m_{H}\simeq 90-130\GeV$.  We have
approximated the loop form factor by just a constant prefactor,
and the numerical value is an approximate upper bound assuming
that we approach $m_H=2\,M_W$ from below. We have taken
$V_{ts}=0.04$ and $m_b=5\GeV$ {for this estimate}. In spite of
the crudeness of the estimate, direct evaluation with programs
FeynArts, FormCalc and LoopTools~\cite{\FeynArts} confirms the
order of magnitude (\ref{estimateBRSM})\,\footnote{To our knowledge, the
  first evaluation of the SM branching ratio was 
performed in~\cite{Eilam:1990zm}, however no detailed analysis of it exists in
the literature. It is natural to clarify this value before jumping to
evaluate the possible non-SM contributions. For the details, see
Ref.\,\cite{Future}.}. On the other hand, if $m_H>2\,M_W$, more
specifically if $m_H>m_t$, it is easy to see that
(\ref{estimateBRSM}) will be suppressed by an additional factor
of $m_b^2/m_H^2$, because the vector boson Higgs decay modes
$H^{SM}\rightarrow W^+\,W^-(Z\,Z)$ will be kinematically
available and become dominant. But at the same time the ratio
${m_H^4}/{m_b^2}$ is replaced by ${m_t^4}/{m_b^2}$. Hence,
\begin{equation}
    BR(H^{SM}\rightarrow b\,\bar{s})
    \sim\left(\frac{|V_{ts}|}{16\pi^2}\right)^2\,\alpha_{W}\
    G_F\,\left(\frac{m_t^4}{m_H^2}\right)\lesssim 10^{-10}\ \ \
    (\,{\rm for}\ m_H>m_t)\,.\label{estimateBRSM2}
\end{equation}
In the numerical evaluation we assumed a mass range where the
ratio $m_H/m_t>1$ is of order one as this provides an upper bound.
In both cases (\ref{estimateBRSM}) and (\ref{estimateBRSM2}) the
branching ratios into bottom quark are much larger than the Higgs
boson FCNC branching ratio into top quark in the
SM\,\cite{Bejar:2003em}. However, even in the case
(\ref{estimateBRSM}) it is still too small to have a chance for
detection in the LHC. It is clear that unless new physics comes to
play the process $H^{SM}\rightarrow b\,\bar{s}$ (and of course
$H^{SM}\rightarrow \bar{b}\,{s}$) will remain virtually invisible.
Nonetheless the result (\ref{estimateBRSM}) is not too far from
being potentially detectable, and one might hope that it should
not be too difficult for the new physics to boost it up to the
observable level.}

{Consider how to estimate the potentially
augmented rates for the MSSM processes (\ref{hFCNC}), if only
within a similarly crude approximation as above. Because of the
strong FCNC gluino couplings mentioned in Section
\ref{sect:introduction} and the $\tb$-enhancement inherent to the
MSSM Yukawa couplings (see Ref.\,\cite{\GuaschSola} for details),
we may expect several orders of magnitude increase of the
branching ratios (\ref{eq:hbs-def}) as compared to the previous SM
result. A naive approach might however go too far. For instance,
one could look at the general structure of the couplings and
venture an enhancement factor typically of order
$(\alpha_s/\alpha_W)^2\,\tan^2\beta\ |\delta_{23}/V_{ts}|^2$, which
for $\delta_{23}\lesssim 1$ and $\tb>30$ could easily rocket the
SM result some $5-6$ orders of magnitude higher, bringing perhaps
one of the MSSM rates (\ref{eq:hbs-def}) to the ``scandalous''
level of $10\%$ or more. But of course only a more elaborated
calculation, assisted by a judicious consideration of the various
experimental restrictions, can provide a reliable result. As we
shall see, a thorough analysis generally disproves the latter
overestimate.}

{The detailed computation of the SUSY-QCD one-loop partial decay
widths $\Ghbs$ in (\ref{eq:hbs-def}) within the MSSM follows
closely that of $\Gamma(t\to ch)$ (see Ref.~\cite{\GuaschSola})}.
{The rather cumbersome analytical expressions will not be listed here
as they are an} straightforward adaptation of those presented {in
the aforementioned references}. {However, there are a few
subtleties that need to be pointed out}.
{One of them is related to the calculation of the total
widths $\Gamma(h\to X)$ for the three Higgs bosons
$h=h^0,H^0,A^0$ in the MSSM}.
{As long as $\Ghbs$ in the numerator of Eq.~(\ref{eq:hbs-def})
is computed at leading order, the denominator has to be computed
also} at leading order}, otherwise an artificial enhancement of
$\Bhbs$ can be generated. For example, including the
next-to-leading (NLO) order QCD corrections to $\Gamma(h\to
b\bar{b})$ reduces the decay width by a significant
amount~\cite{\hbbQCD}. {Then, to be consistent,} the NLO
(two-loop) contributions to $\Ghbs$ should also be included.
{Similarly}, the one-loop SUSY-QCD corrections to $\Gamma(h\to
b\bar{b})$ can be very large and negative~\cite{Coarasa:1996yg},
which would enhance $\Bhbs$. At the same time these corrections
also contribute to $\Ghbs$, such that contributions to the
numerator and denominator of Eq.~(\ref{eq:hbs-def}) compensate
(at least partially) each other. Therefore the same order of
perturbation theory must be used in both partial decay widths
entering the observable $\Bhbs$ to obtain a consistent result.
{By the same token, using running masses in the numerator
of (\ref{eq:hbs-def}) is mandatory, if they are used in the
denominator.}
{Last, but not least, consistency with the experimental
bounds on related observables should also be taken into account.
In this respect an essential role is played by the constraints on
the FCNC couplings from the measured value of $\Bbsg$. They must
be included in this kind of analysis, if we aim at a realistic
estimate of the maximal rates expected for the FCNC processes
(\ref{hFCNC}) in the MSSM. In our calculation we have used the
full one-loop MSSM contributions to $\Bbsg$ as given
in~\cite{Urban}}\,\footnote{Ref.\cite{Urban} contains a partial two-loop
  computation of $\Bbsg$, but only the one-loop contributions have been
used for the present work.}.

Let us now summarize the {conditions under which we have
performed the computation and the approximations and assumptions
made in the present analysis}:
\begin{itemize}
\item We include the full one-loop SUSY-QCD contributions to the partial
  decay widths $\Ghbs$ in (\ref{eq:hbs-def}).
\item We assume that FCNC mixing terms appear only in the
  {LH-chiral} sector of the squark mixing matrix. This is
  the most natural assumption, and, moreover, it was proven in
  Ref.~\cite{\GuaschSola} that the presence of FCNC terms in the
  {RH-chiral sector} enhances the partial widths by a factor two
  at most -- not an order of magnitude.
\item The Higgs bosons total decay widths $\Gamma(h\to X)$ {are} computed
  at leading order, including all the relevant channels: $\Gamma(h\to
  f\bar{f},ZZ,W^+W^-,gg)$. The off-shell decays
  $\Gamma(h\to ZZ^*,W^{\pm}W^{\mp*})$ have also been
  included. {The one-loop decay rate $\Gamma(h\to gg)$
  has been taken from~\cite{Spira:1995rr} {and the off-shell decay
  partial widths have been computed explicitly and found perfect agreement with the old
  literature on the subject\,\cite{Keung:1984hn}.}
  We have verified that some of the aforementioned higher order decays
   are essential to consistently compute the
  total decay width of $\Gamma(h^0\to X)$ in certain regions of the parameter
  space where the maximization procedure probes domains in which some
(usually leading) two-body processes become {greatly} diminished.
  We have checked that our
  implementation of
  the various Higgs boson decay rates is consistent with the results
  of \texttt{HDECAY}~\cite{hdecay}. However, care must be
  exercised if using the full-fledged result from \texttt{HDECAY}.
  For example, it would be inconsistent,
  and numerically significant, to compute the
  total widths $\Gamma(h\rightarrow X)$ with this program and at the same time
  to compute the SUSY-QCD one-loop partial widths $\Ghbs$ without including the leading
  conventional QCD effects through e.g. the running quark masses.}

\item The Higgs sector parameters (masses and CP-even mixing angle
  $\alpha$) {have} been treated using the leading $\mt$ and $\mb\tb$
  approximation  to the one-loop result~\cite{\Dabels}. For comparison, we also
  perform the analysis using the tree-level approximation.
\item We include the {constraints} on the {MSSM parameter space} from
  $\Bbsg$. {We adopt $\Bbsg=(2.1-4.5)\times 10^{-4}$ as the experimentally allowed
  range within three standard deviations}~\cite{Hagiwara:2002fs}. {Only the
SUSY-QCD contributions
  induced from tree-level FCNCs are considered in the present work.}

\end{itemize}
Running quark masses ($m_q(Q)$) and strong coupling constants
($\alpha_s(Q)$) are used throughout. More details are given
below, as necessary.

\section{Full one-loop SUSY-QCD calculation: Numerical analysis}
\label{sect:numerical}

Given the setup described in Section \ref{sect:partialwidths}, we
have performed a {systematic scan} of the MSSM parameter
space with the following restrictions:
\begin{equation}
\begin{array}{rcl}
\delta_{23}&<&10^{-0.09} \simeq 0.81  \\
A_b&=&-1500 \cdots 1500 \GeV\\
\mu&=&-1000 \cdots 1000 \GeV\\
m_{\squark}&=&150 \cdots 1000 \GeV
\end{array}
\label{eq:scan-parameters}
\end{equation}
and the following fixed parameters:
\begin{equation}
\begin{array}{rcl}
\tan\beta&=&50 \\
m_{\sbottom_L}&=&m_{\sbottom_R}=m_{\stopp_R}=\mg=m_{\squark}\\
A_t&=&-300 \GeV\,\,.
\end{array}
\label{eq:scan-fixed}
\end{equation}
Here $m_{\sbottom_{L,R}}$ are the  left-{chiral} and
right-chiral bottom-squark soft-SUSY-breaking mass parameters, and
$m_{\squark}$ is a common mass for the strange- and down-squark
left- and right-chiral  soft SUSY-breaking mass parameters.
{Following the same notation as in~\cite{Guasch:1999jp}, the
parameter $\delta_{23}$  represents the mixing between the second
and third generation squarks. Let us recall its definition:}
\begin{equation}\label{delta23}
\delta_{23}\equiv \frac{m^2_{\sbottom_L
\sstrange_L}}{m_{\sbottom_L} m_{\sstrange_L}}\,,
\end{equation}
$m^2_{\sbottom_L \sstrange_L}$ {being} the non-diagonal
term in the squark mass matrix squared mixing the second and third
generation left-chiral squarks. {The parameter
$\delta_{23}$ is a fundamental parameter in our analysis as it
determines the strength of the tree-level FCNC interactions
induced by the supersymmetric strong interactions, which are then
transferred to the loop diagrams of the Higgs boson FCNC decays
(\ref{hFCNC}).}

\FIGURE[pt]{%\begin{figure}
\begin{tabular}{cc}
\resizebox{!}{6cm}{\includegraphics{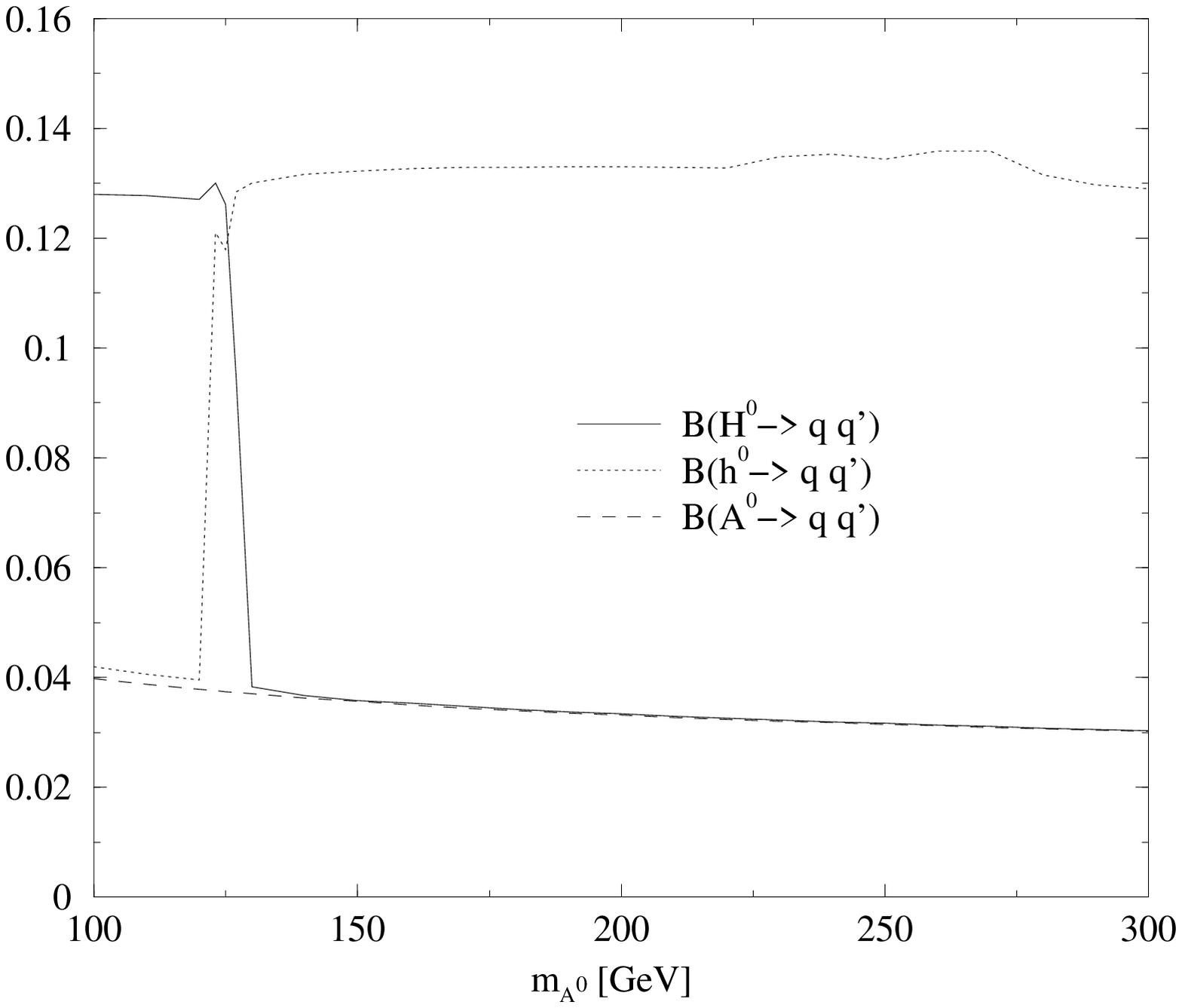}} &
\resizebox{!}{6cm}{\includegraphics{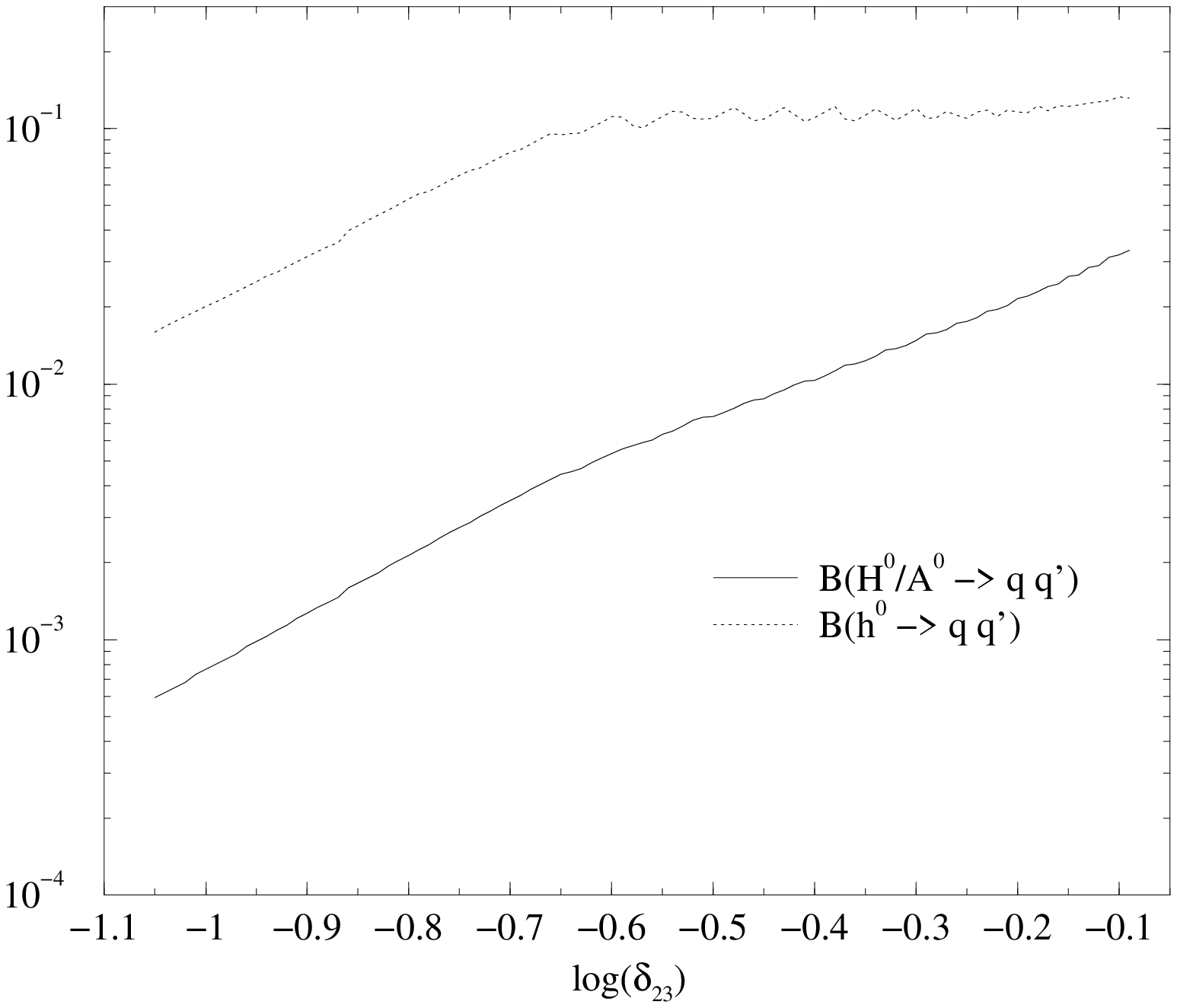}} \\
(a) & (b)
\end{tabular}
\caption{{Maximum SUSY-QCD contributions to
$B(h\rightarrow q\,q')$, Eq.\,(\ref{eq:hbs-def}), as a
  function of} \textbf{a)} $\ma$ and \textbf{b)} $\delta_{23}$ for
  $\ma=200\GeV$.}\label{fig:maximfull}
}%\end{figure}

\TABLE[pt]{%\begin{table}
\centerline{\begin{tabular}{|c||c|c|c|} \hline Particle &  $H^0$
& $h^0$ & $A^0$ \\\hline\hline 
\Bhbs &  $3.3\times 10^{-2}$ & $1.3\times 10^{-1}$ & $3.3\times
10^{-2}$\\
\hline $\Gamma(h\to
X)$ & $11.0 \GeV$ & $1.6\times 10^{-3} \GeV$ & $11.3 \GeV$
\\\hline $\delta_{23}$ & $10^{-0.09}$& $10^{-0.1}$ &
$10^{-0.09}$\\\hline $m_{\squark}$ & $975 \GeV$ &  $975 \GeV$ &
$975 \GeV$\\\hline $A_b$ & $1500 \GeV$ & $730 \GeV$ &
$1290\GeV$\\\hline $\mu$ & $980 \GeV$ & $1000 \GeV$ & $980 \GeV$
\\\hline \Bbsg &  $4.42\times 10^{-4}$ &  $4.23\times 10^{-4}$
&$4.50\times 10^{-4}$ \\\hline
\end{tabular}}
\caption{Maximum values of $\Bhbs$ and corresponding SUSY
parameters for
  $\ma=200\GeV$.}\label{tab:maxim1}
}%\end{table}

{The result of the scan is depicted in Fig.~\ref{fig:maximfull}.
To be specific: Fig.~\ref{fig:maximfull}a shows  the maximum value
$\Bhbsmax$ of the FCNC decay rate (\ref{eq:hbs-def}) under study
as a function of $\ma$; Fig.~\ref{fig:maximfull}b displays
$\Bhbsmax$ as a function of the mixing parameter $\delta_{23}$ for
$\ma=200\GeV$}. Looking at Fig.~\ref{fig:maximfull} three facts
strike the eye immediately :
{i) the maximum is huge ($ 13\%$!) for a FCNC rate, actually it is as
big as initially guessed from the rough estimates made in Section
\ref{sect:partialwidths}}; ii) very large values of $\delta_{23}$
are allowed; iii) the maximum rate is independent of the
pseudo-scalar Higgs boson mass $\mA$.
{We will now analyze facts ii) and iii) in turn,
and will establish their incidence on fact i)}. For further
reference, in Table~\ref{tab:maxim1} we show the numerical values
of $\Bhbsmax$ together with the parameters which maximize the
rates for $\ma=200\GeV$.

\FIGURE[pt]{%\begin{figure}[tb]
\begin{tabular}{cc}
\resizebox{!}{6cm}{\includegraphics{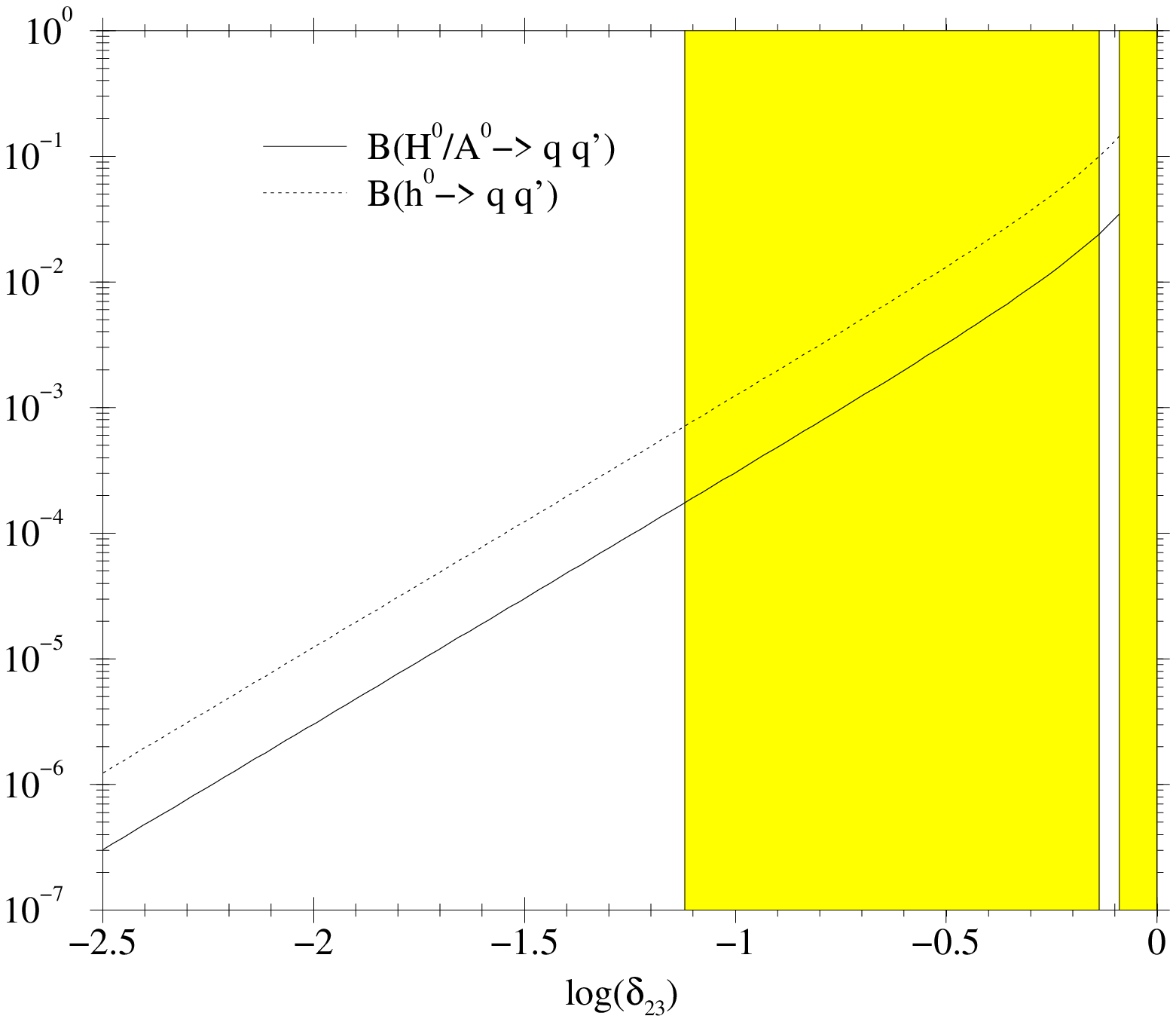}} &
\resizebox{!}{6cm}{\includegraphics{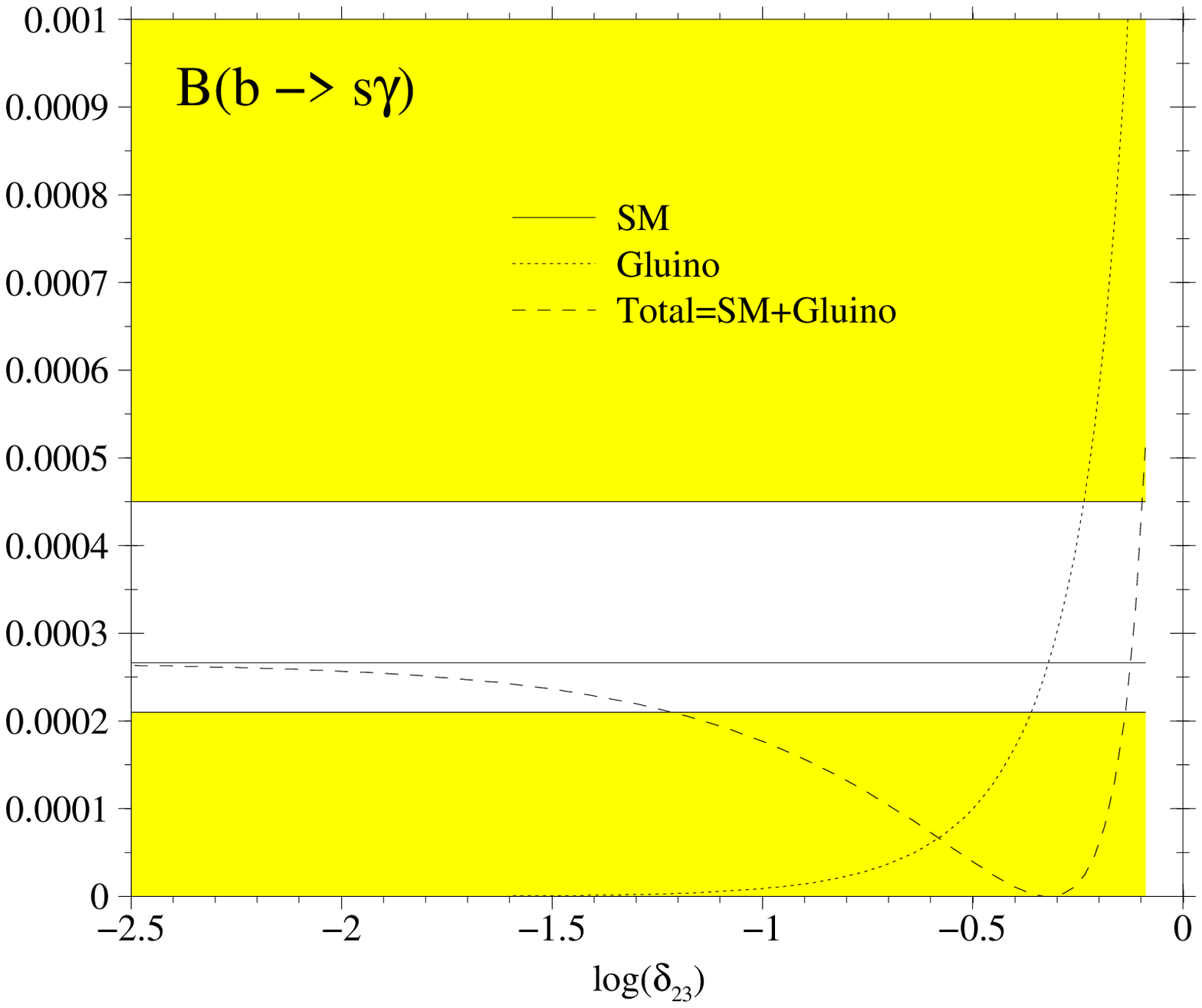}} \\
(a) & (b)
\end{tabular}
\caption{$\Bhbs$ and $\Bbsg$ as a function of $\delta_{23}$ for
the
  parameters that maximize $\Bhzbs$ in
  Table~\protect{\ref{tab:maxim1}}. The shaded region is excluded experimentally.}\label{fig:d23bsg}
}%\end{figure}

\FIGURE[pt]{%\begin{figure}
\begin{tabular}{cc}
\resizebox{!}{6cm}{\includegraphics{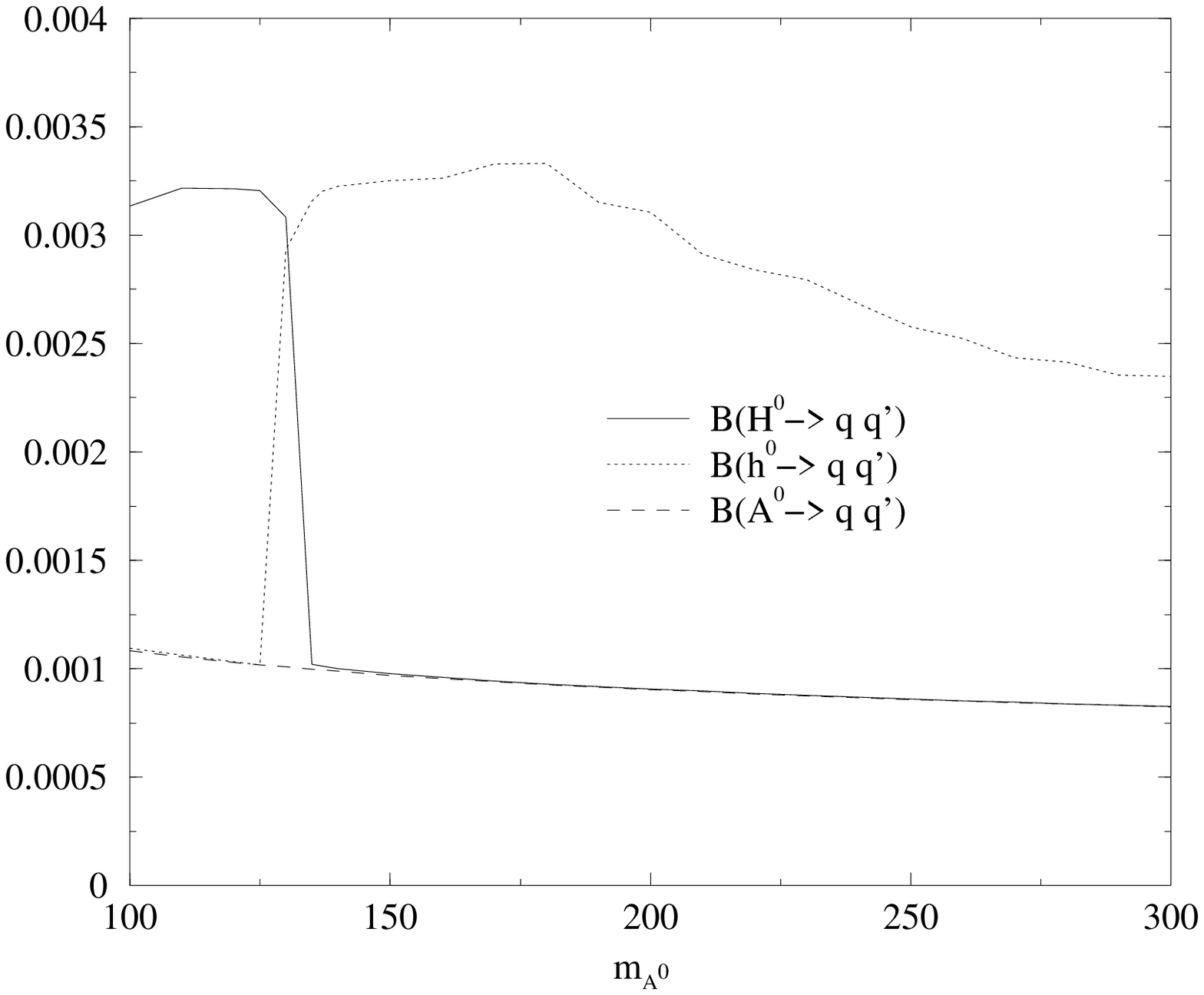}} &
\resizebox{!}{6cm}{\includegraphics{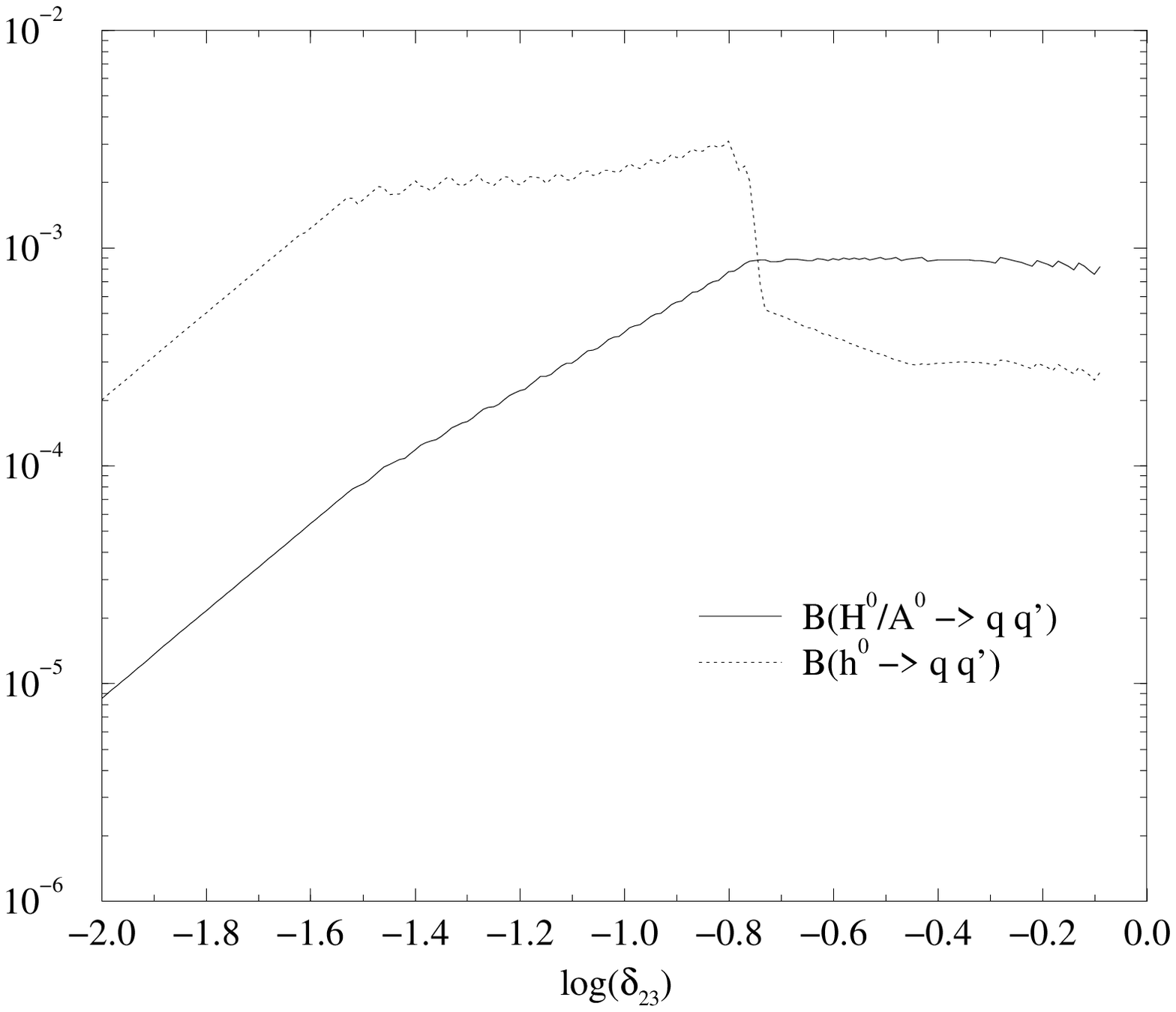}} \\
(a) & (b)
\end{tabular}
\caption{Maximum value of the SUSY-QCD contributions to $\Bhbs$
as a
  function of \textbf{a)} $\ma$ and \textbf{b)} $\delta_{23}$ for
  $\ma=200\GeV$, for the scenario excluding the \textit{window} regions.\label{fig:maximnowindow}}
}%\end{figure}

\TABLE[pt]{%\begin{table}[tb]
\centerline{\begin{tabular}{|c||c|c|c|} \hline Particle &  $H^0$
& $h^0$ & $A^0$ \\
\hline\hline \Bhbs &  
$9.1\times 10^{-4}$ & $3.1\times 10^{-3}$ & $9.1\times 10^{-4}$\\
\hline $\Gamma(h\to
X)$ & $11.2 \GeV$ & $1.4\times 10^{-3} \GeV$ & $11.3 \GeV$
\\\hline $\delta_{23}$ & $10^{-0.43}$& $10^{-0.8}$ &
$10^{-0.43}$\\\hline $m_{\squark}$ & $1000 \GeV$ &  $975 \GeV$ &
$1000 \GeV$\\\hline $A_b$ & $-1500 \GeV$ & $-1500 \GeV$ & $-1500
\GeV$\\\hline $\mu$ & $-460 \GeV$ & $-1000 \GeV$ & $-460 \GeV$
\\\hline \Bbsg &  $4.49\times 10^{-4}$ &  $4.48\times 10^{-4}$
&$4.49\times 10^{-4}$ \\\hline
\end{tabular}}
\caption{Maximum values of $\Bhbs$ and corresponding SUSY
parameters for
  $\ma=200\GeV$ excluding the \textit{window} region.\label{tab:maximnowindow}}
}%\end{table}

One would expect that a large value of $\delta_{23}$ should
induce a large gluino contribution to $\Bbsg$. In fact it does!
However our automatic scanning process picks up the corners of
parameter space where the gluino contribution alone is much
larger than the SM contribution, but opposite in sign, such that
both contributions destroy themselves partially leaving a result
in accordance with the experimental {constraints}}. We
examine this behaviour in Fig.~\ref{fig:d23bsg}, where we show
the {values} of $\Bhbs$ together with $\Bbsg$ as a
function of $\delta_{23}$ for the parameters which maximize {the
FCNC rate of the lightest CP-even state $h^0$} in
Table~\ref{tab:maxim1}. 
We see that, for small values of
$\delta_{23}$, the gluino contribution to $\Bbsg$ is small, and
the total $\Bbsg$ prediction is close to the SM
expectation. {In contrast,  as $\delta_{23}$ steadily grows, $\Bbsg$
decreases fast (meaning a dramatic cancellation between the two
contributions) until reaching a point where $\Bbsg=0$}. From
there on it starts to grow with a large slope, and in its race
eventually crosses the allowed $\Bbsg$ region.
{The crossing is very fast, and so rather ephemeral in the
$\delta_{23}$ variable, and it leads to the appearance of a narrow
allowed \textit{window}
{at} large $\delta_{23}$ values, see  Fig.~\ref{fig:d23bsg}a}.
 {We would regard the choice
of this window as a fine-tuning of parameters, hence unnatural}.
For this reason we reexamine the $\Bhbs$ ratio by performing
{a new scan of the MSSM parameter space in which we
exclude the fine-tuned (or \textit{window}) region}. The result
for $\mA=200\GeV$ can be seen in Table~\ref{tab:maximnowindow} and
Fig.~\ref{fig:maximnowindow}. This time we see that the maximum
values of $\Bhbs$ are obtained for much lower values of
$\delta_{23}$, and the maximum rates have decreased more than one
order of magnitude with respect to Table~\ref{tab:maxim1},
{reaching the level of few per mil}. These FCNC rates can still
be regarded as fantastically large. Had we included the
{SUSY-EW}
  contributions to $\Bbsg$, further cancellations might have
  occurred between the SUSY-EW and the SUSY-QCD amplitudes. Even more: since each
    contribution depends on a separate set of parameters, one would be able to
    find a set of parameters in the SUSY-EW sector which creates an
    amplitude that compensates the SUSY-QCD contributions for almost any
    point of the SUSY-QCD parameter
    space{\,\cite{Future}}. But of course this would be only at the price of
    performing some
    fine tuning, which is not the approach we want to follow here.

    On the other hand further
      contributions to $\Bbsg$ might exist. In the most general MSSM, flavor-changing
      interactions  for the right-chiral squarks
      ($\delta_{23RR}$), and mixing left- and right-chiral squarks
      ($\delta_{23LR}$) can be introduced. The latter can produce
      significant contributions to 
    $\Bbsg$, changing the allowed parameter space. The
    introduction of $\delta_{23LR}$ can produce two possible outcomes:
    First, in certain regions of the parameter space, the contributions
    of $\delta_{23LR}$ and $\delta_{23}$ are of the same sign, enhancing
    each other. In this situation, the maximum allowed value of
    $\delta_{23}$ is obtained for $\delta_{23LR}=0$. Second, in other
    regions of the parameter space the two contributions would
    compensate each other, producing an overall value of $\Bbsg$ in
    accordance with experimental constraints, even though each
    contribution would be much larger. Again, we would regard these
    compensations as unnatural, and would discard that region of the
    parameter space. In the following we will require that the SUSY-QCD
    contributions induced by $\delta_{23}$ do not compensate the SM
    ones to give an acceptable 
    value of $\Bbsg$; this is equivalent to 
    the condition that the SUSY-QCD amplitude  represents a
    small contribution to the total $\Bbsg$ value, and is therefore
    independent of the inclusion of the other 
    contributions (SUSY-EW,
    $\delta_{23LR}$).\footnote{The analysis of
        Ref.~\cite{Demir:2003bv} follows the opposite approach, that is:
      to find the fine-tuning conditions imposed by low energy data that
    allow for the largest possible value of the FCNC parameters.}

%\begin{figure}[tb]
\FIGURE[pt]{\begin{tabular}{cc}
\resizebox{!}{6cm}{\includegraphics{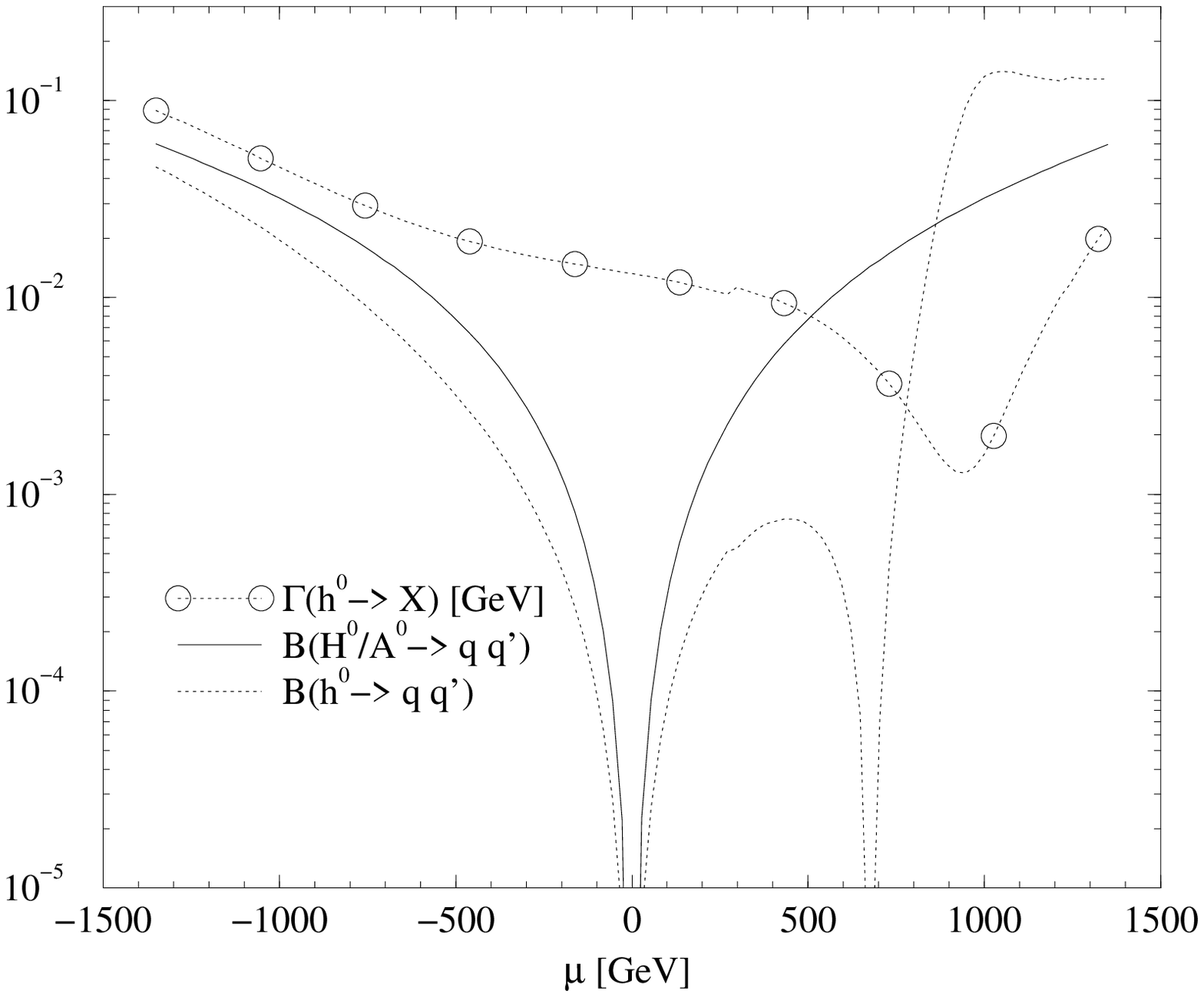}} &
\resizebox{!}{6cm}{\includegraphics{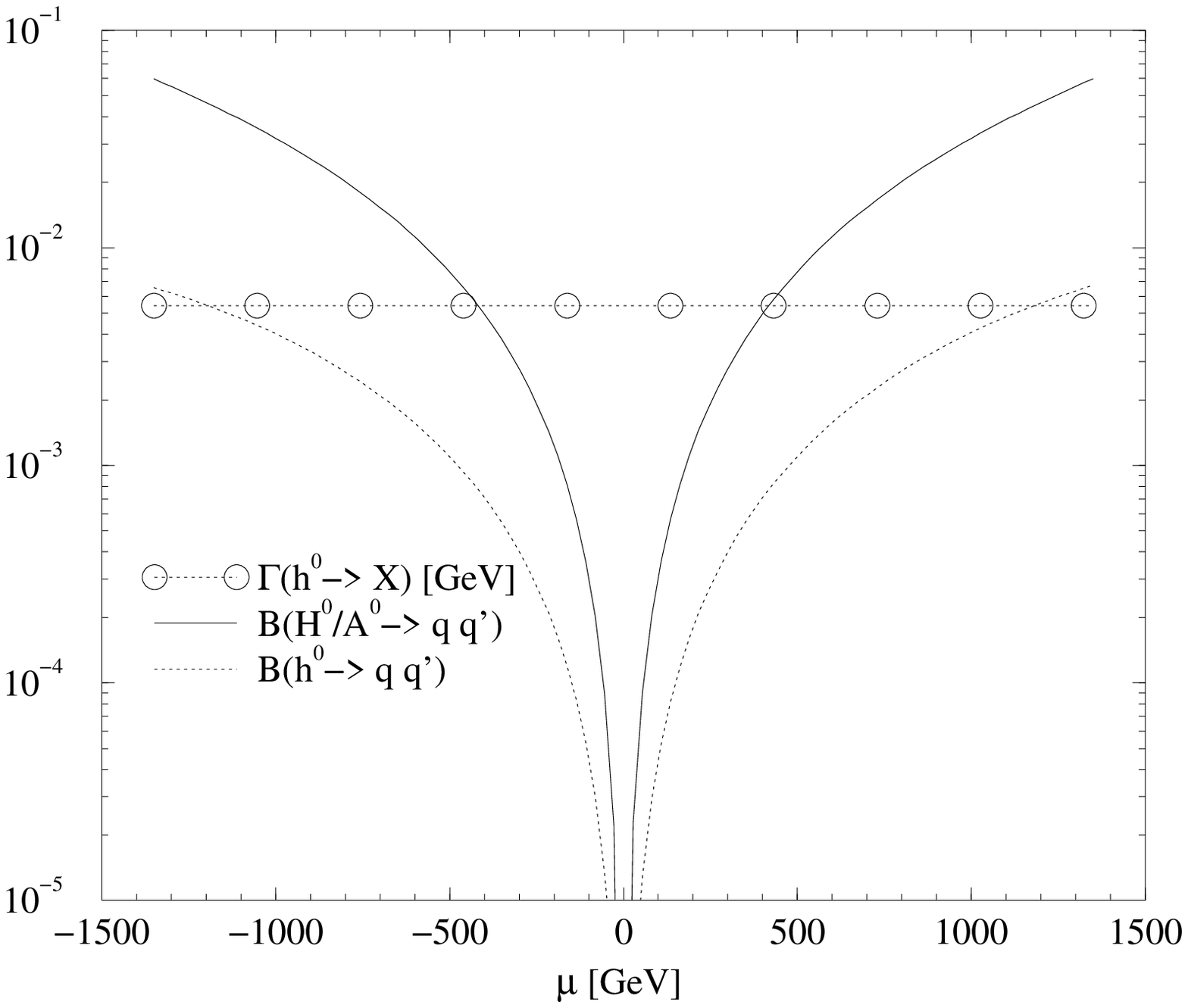}} \\
(a) & (b) 
\end{tabular}
\caption{$\Bhbs$ and $\Gamma(h^0\to X)$ (in\GeV) as a function of $\mu$ for
  \textbf{a)} one-loop $\alpha$ angle; \textbf{b)} tree-level $\alpha$
  angle, and for the
  parameters that maximize $\Bhzbs$ in
  Table~\ref{tab:maxim1}.
  The $H^0$ and $A^0$ curves coincide. The $\Bbsg$ {constraint} is
  not shown.\label{fig:alphamu}}
}%\end{figure}

We turn now our view to the second fact, namely the independence
of the maximum rates with respect to $\ma$. {We will show
that it also plays a central role as to the enhancement of
$\Bhbs$}. Actually, a good hint is given by the small values of
the lightest Higgs boson decay width in Tables~\ref{tab:maxim1}
and~\ref{tab:maximnowindow}, $\Gamma(h^0\to X)\sim 2\times
10^{-3}\GeV$. The maximization process of $\Bhzbs$ does not only
find the parameters for which $\Ghzbs$ is maximum, but also the
parameters for which $\Gamma(h^0\to X)$ is minimum. Specifically,
since $\Gamma(h^0 \to b\bar{b})$ is the dominant decay decay
channel of $h^0$ for large $\tb$, the maximum of $\Bhzbs$ is
produced in the parameter range of the so-called \textit{small
$\alpha_{eff}$
  scenario}~\cite{Carena:2002qg}, that is, a parameter range where the
radiative corrections make the CP-even Higgs boson mixing angle
$\alpha$ vanish (or very small), such that the leading partial
decay width $\Gamma(h^0\to b\bar{b})$ is strongly suppressed. The
consequences of this scenario have been extensively studied in
Ref.~\cite{Carena:1999bh}. {As advertised in Section
\ref{sect:partialwidths}, the possibility that the maximization
process explores these regions of the parameter space is the
reason why the leading higher order decay channels, and also the
leading three-body decay modes have to be taken into account in
the computation of the total width.}

{In Fig.~\ref{fig:alphamu} we plot the value of the various
branching ratios $\Bhbs$ and of the total width of the lightest
CP-even Higgs boson, $\Gamma(h^0\to X)$, as a function of the
higgsino mass parameter $\mu$}, the rest of the parameters being
{those of the third column of Table~\ref{tab:maxim1},
i.e. the ones that maximize the branching ratio $\Bhzbs$}.
Fig.~\ref{fig:alphamu}a shows that $\Gamma(h^0\to X)$ has a deep
minimum in the range of $\mu$ corresponding to the maximum of
$\Bhzbs$, {which reaches the level of a few percent}. If, instead
of using the radiatively corrected $\alpha$ value we use the
tree-level expression, we obtain the result shown in
Fig.~\ref{fig:alphamu}b. Here the total decay width of the Higgs
boson is independent of $\mu$, and $\Bhbs$ does not show any
peak. {Actually in this case the branching ratio for $h^0$
becomes smaller than that of $H^0$ and $A^0$ for all $\mu$.} The
maximization procedure in Fig.~\ref{fig:maximfull} selects for
each value of $\ma$ the MSSM parameters corresponding to the
small $\alpha_{eff}$ scenario for that specific value of $\ma$.
Of course, this discussion regarding the $h^0$ channels for large
values of $\ma$ has a correspondence {with} the $H^0$ channel for
low values\footnote{{Large or low values here means
$\ma>m_{h^0}^{\rm max}$ or $\ma<m_{h^0}^{\rm max}$, i.e. above or
below the maximum possible value for the mass of the lightest
Higgs boson $h^0$, respectively}.} of $\ma$.

{As indicated in Section \ref{sect:partialwidths},
we have used a one-loop approximation for the Higgs
sector}~\cite{\Dabels}, instead of the more sophisticated
complete two-loop result present in the
literature~\cite{Carena:2000dp,Espinosa:2000df}. However, we
should stress that the exact MSSM parameters at which the small
$\alpha_{eff}$ scenario is realized are not important for the
sake of the present analysis. All that matters is that some
portion of the parameter space exists, for which $\Gamma(h^0\to
b\bar{b})$ is strongly suppressed, {but $\Ghzbs$ is not}.

\TABLE[pt]{%\begin{table}
\centerline{\begin{tabular}{|c||c|c|c|} \hline Particle &  $H^0$
& $h^0$ & $A^0$ \\
\hline\hline \Bhbs & 
 $9.0\times 10^{-4}$ & $1.3\times 10^{-4}$ & $9.0\times 10^{-4}$\\
\hline $\Gamma(h\to
X)$ & $11.3 \GeV$ & $5.4\times 10^{-3} \GeV$ & $11.3 \GeV$
\\\hline $\delta_{23}$ & $10^{-0.43}$& $10^{-0.28}$ &
$10^{-0.43}$\\\hline $m_{\squark}$ & $1000 \GeV$ &  $1000 \GeV$ &
$1000 \GeV$\\\hline $A_b$ & $-1500 \GeV$ & $-1500 \GeV$ & $-1500
\GeV$\\\hline $\mu$ & $-460 \GeV$ & $-310 \GeV$ & $-460 \GeV$
\\\hline \Bbsg &  $4.49\times 10^{-4}$ &  $4.50\times 10^{-4}$
&$4.49\times 10^{-4}$ \\\hline
\end{tabular}}
\caption{Maximum values of $\Bhbs$ and corresponding SUSY
parameters for
  $\ma=200\GeV$, using the tree-level expressions for the Higgs sector,
  and excluding the \textit{window} region.\label{tab:maximnowindowtree}}
}%\end{table}

\FIGURE[pt]{%\begin{figure}
\begin{tabular}{cc}
\resizebox{!}{6cm}{\includegraphics{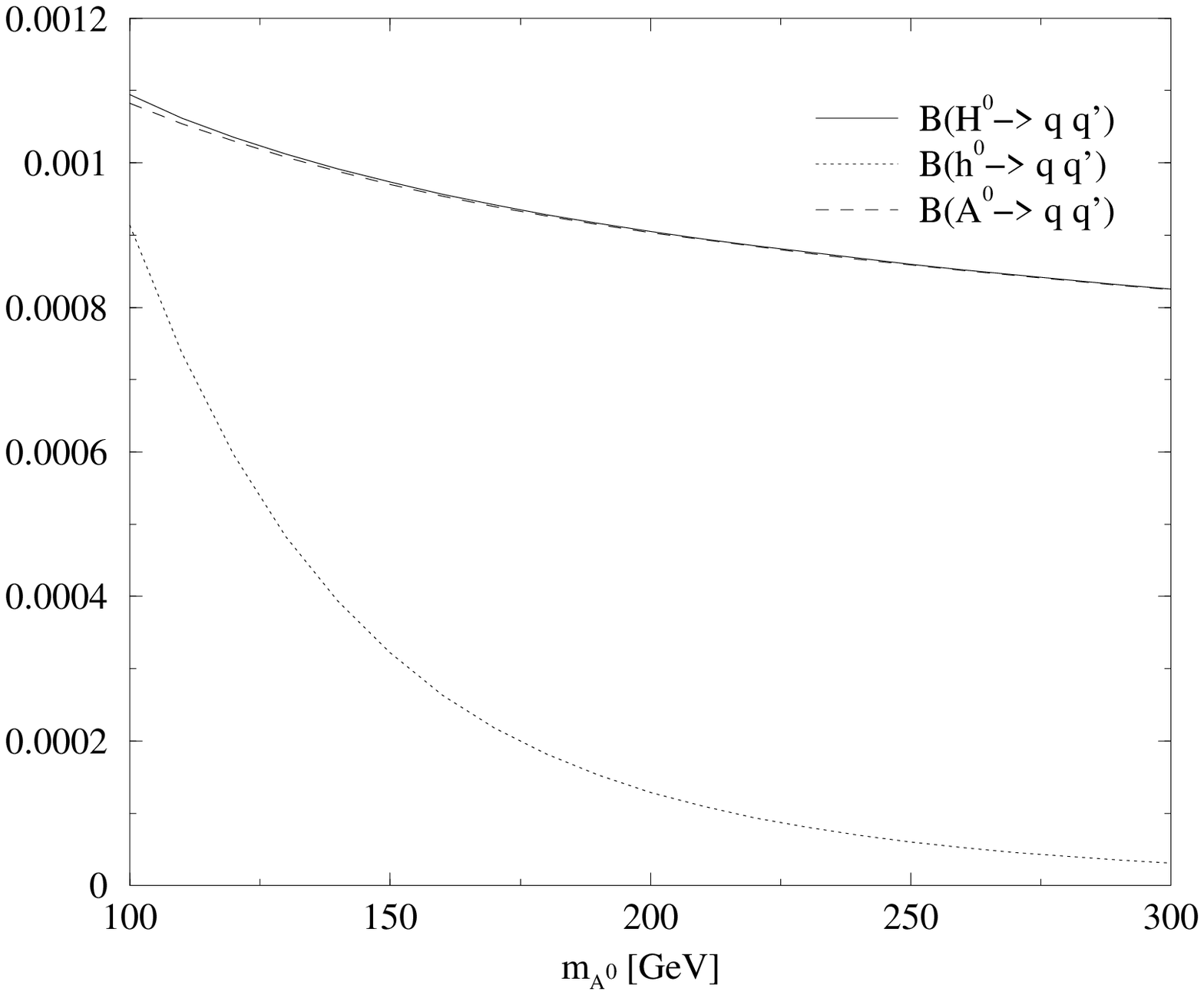}} &
\resizebox{!}{6cm}{\includegraphics{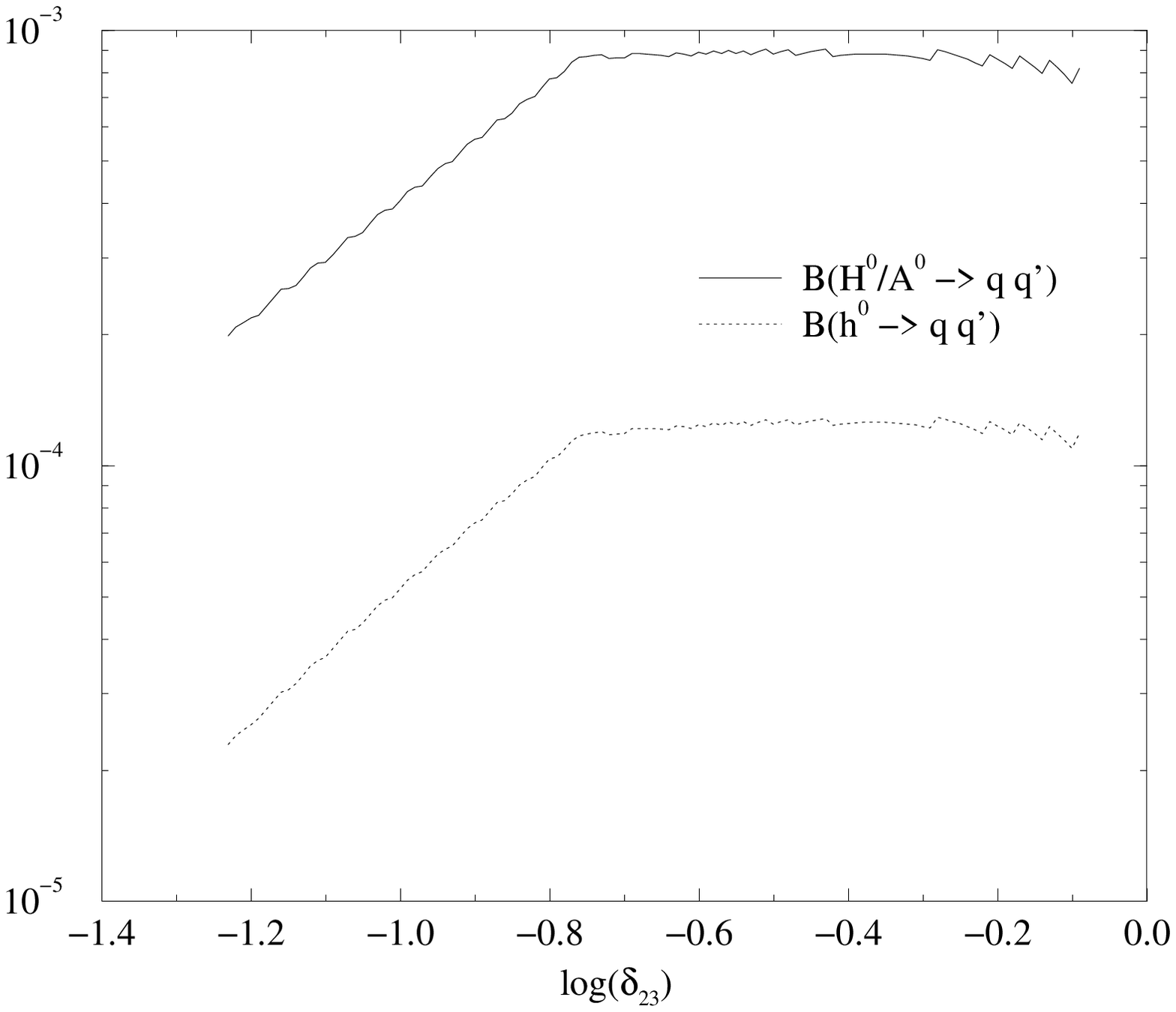}} \\
(a) & (b)
\end{tabular}
\caption{Maximum value of the SUSY-QCD contributions to $\Bhbs$
as a
  function of \textbf{a)} $\ma$ and \textbf{b)} $\delta_{23}$, for
  $\ma=200\GeV$ and for the scenario excluding the \textit{window} region
  and using the tree-level expressions for the Higgs sector
  parameters.\label{fig:maximnowindowtree}}
}%\end{figure}

{To compare the maximum value of $\Bhzbs$ obtained with and
without the small $\alpha_{eff}$ scenario, we have performed the
maximization procedure using the tree-level expressions for the
{Higgs sector parameters}}. The result is shown {in}
Table~\ref{tab:maximnowindowtree} and
Fig.~\ref{fig:maximnowindowtree}. {In this case $\Bhzbs$ is
reduced by a sizeable factor of $\gtrsim 20$ with respect {to}
Table~\ref{tab:maximnowindow}, whereby the $h^0$ rate descends
about an order of magnitude below that of the $H^0/A^0$ channels
which remain basically unchanged}. Notice also that
$\Gamma(h^0\to X)$ is larger than in previous tables. In spite of
the reduction, achieving a FCNC ratio  $\Bhzbs\sim 1.3 \times
10^{-4}$ is a remarkable result, {three orders of magnitude}
larger than the maximum SM rate (\ref{estimateBRSM}), and only
one order of magnitude below the rare decay} $B(h^0\to
\gamma\gamma)\sim 10^{-3}$.
{Also worth noticing
in Fig.~\ref{fig:maximnowindowtree}b (and
Fig.~\ref{fig:maximnowindow}b) is the fact that $\Bhbsmax$ is
essentially flat in $\delta_{23}$ in the upper range {down to}
$\delta_{23}\sim10^{-0.8}\simeq 0.16$}. The reason lies in the
correlation between $\Bhbs$ and $\Bbsg$. In order to comply with
the (non-fine-tuned) value of $\Bbsg$ for large $\delta_{23}$, the
absolute value of the $\mu$ parameter must be small. When
$\delta_{23}$ decreases, $|\mu|$ can grow to larger values,
leaving the overall {maximum rates} $\Bhbsmax$
effectively unchanged (see Eq.~(\ref{eq:approxleading}) below).

The maximization process selects a squark mass scale in the
vicinity of the maximum values used in the scanning procedure. We
should point out, however, that the same order of magnitude for
$\Bhbs$ could be obtained with a much lower squark {mass}
scale. {In this case} the lighter squark masses induce a
much larger $\Bbsg$ value, and $\delta_{23}$ is much more
constrained. For example,
if we perform a {scan} in the parameter
space~(\ref{eq:scan-parameters}), but fixing the squark mass scale to be
$m_{\squark}<500\GeV$, we obtain the following values for the {maximal} branching
ratios for $\ma=200\GeV$:
\begin{equation}
\Bhzbsmax=1.4\times 10^{-5}\   , \ 
\BHAbsmax=9.2\times 10^{-5}\   ,
\label{eq:brlowscale}
\end{equation}
with $\delta_{23}\sim 10^{-0.6}$, $\mu\sim-110\GeV$, and we have limited
ourselves to the scenario avoiding the 
\textit{window} regions and using the tree-level expression for the Higgs
sector parameters. These numbers have to be compared with
Table~\ref{tab:maximnowindowtree}. 

The reason behind this \textit{scale independence} admits an explanation
in terms of an effective Lagrangian approach~\cite{Future}, in which one
can estimate the leading effective coupling to behave approximately as:
\begin{eqnarray}
g_{h b\bar{s}}   &\simeq& \frac{g\mb}{\sqrt{2}\mw \cos\beta} \frac{2
    \alpha_s}{3 \pi}  \delta_{23} \frac{-\mu \,
    \mg}{M_{SUSY}^2} \left\{ \begin{array}{cc}
 \sin(\beta-\alpha)& (H^0)\\
    \cos(\beta-\alpha) & (h^0)\\
 \sin(2\beta)& (A^0)
  \end{array}\right.\ \ .
\label{eq:approxleading}
\end{eqnarray}
Aside from {ensuring (at least) a} partial SUSY scale
independence of the leading terms, this expression also shows
that $\Bhbs$ has a weak dependence on the soft-SUSY-breaking
trilinear coupling $A_b$. 
The observed situation is
similar to the flavor-conserving $hb\bar{b}$ interactions, where
the cancellation of the $A_b$ terms at leading order has been
recently proven~\cite{SpiraGuasch}. It also shows that the
leading {non-decoupling SUSY contributions to $\Ghzbs$
eventually fade out as the decoupling limit of the Higgs sector is
approached: $\cos(\beta-\alpha)\to 0$}. We have found (using the
tree-level expression for $\alpha$) that the non-leading
({SUSY-decoupling}) contributions to $\Ghzbs$ dominate for
$\mA\gsim 450\GeV$, inducing a value $\Ghzbsmax\sim 1.2\times
10^{-5}$, with $\delta_{23}\sim 10^{-1}$, $\mu\sim1000\GeV$. Full
details on the effective Lagrangian approach, {and its
application to further refine these calculations}, will be given
in a forthcoming publication~\cite{Future}.

\FIGURE[pt]{%\begin{figure}
\begin{tabular}{cc}
\resizebox{!}{6cm}{\includegraphics{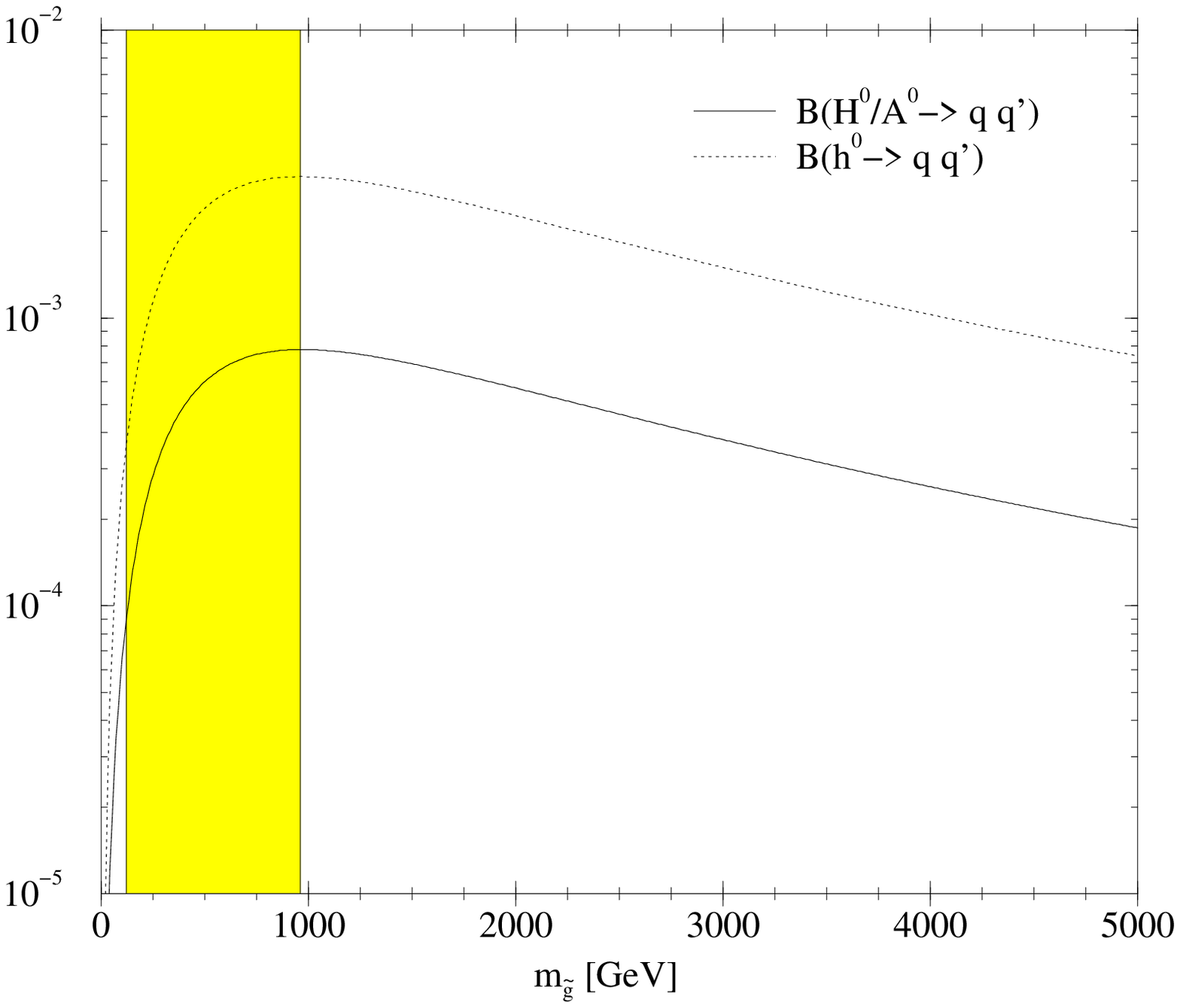}} &
\resizebox{!}{6cm}{\includegraphics{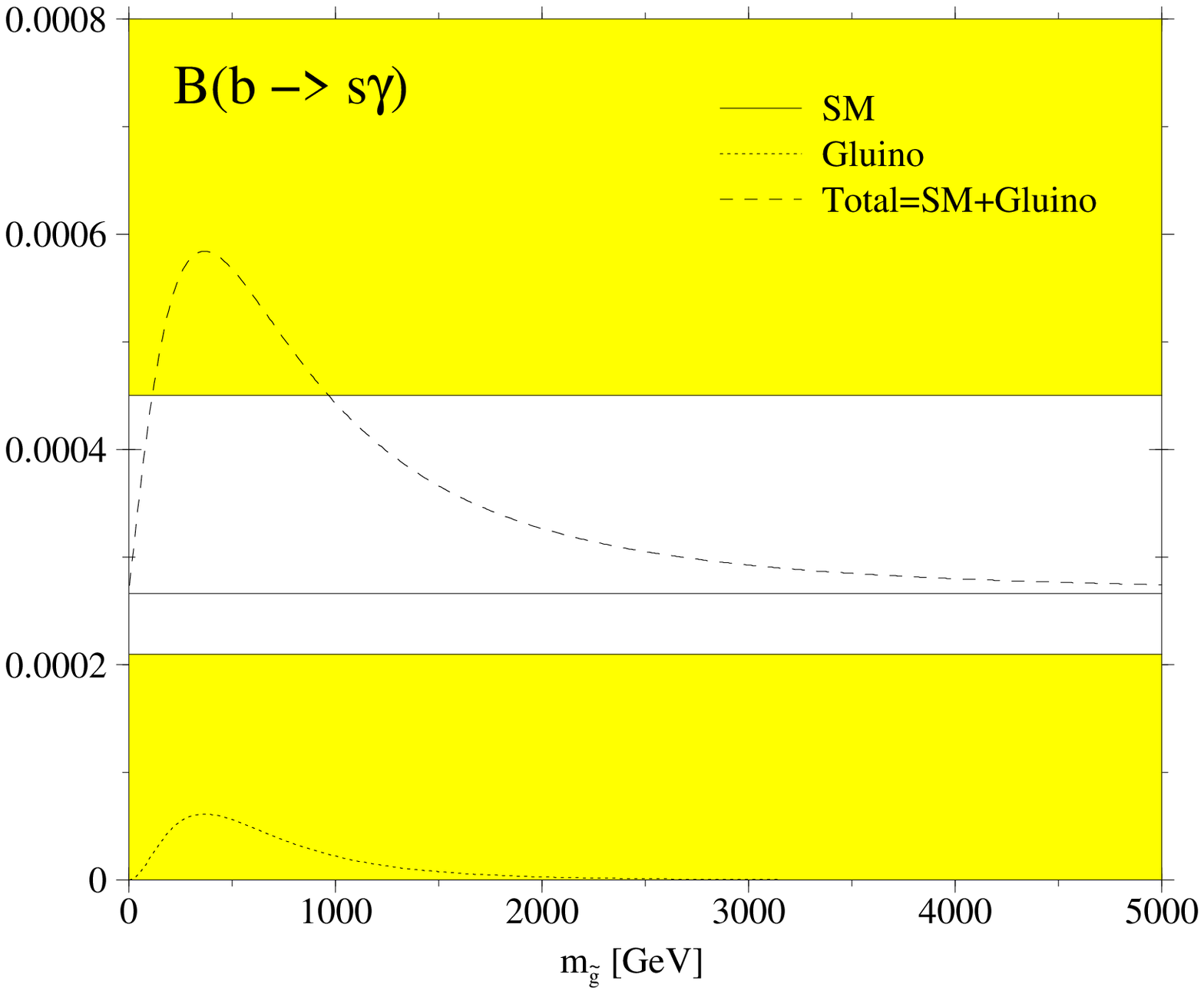}} \\
(a) & (b)
\end{tabular}
\caption{$\Bhbs$ and $\Bbsg$ as a function of $\mg$ for the
  parameters 
  that maximize $B(h^0\to b\bar{s})$ excluding the window
  region {(see third
  column of Table~\protect{\ref{tab:maximnowindow}}).} The shaded region is excluded experimentally.\label{fig:mg}}
}%\end{figure}

\FIGURE[pt]{%\begin{figure}
\begin{tabular}{cc}
\resizebox{!}{6cm}{\includegraphics{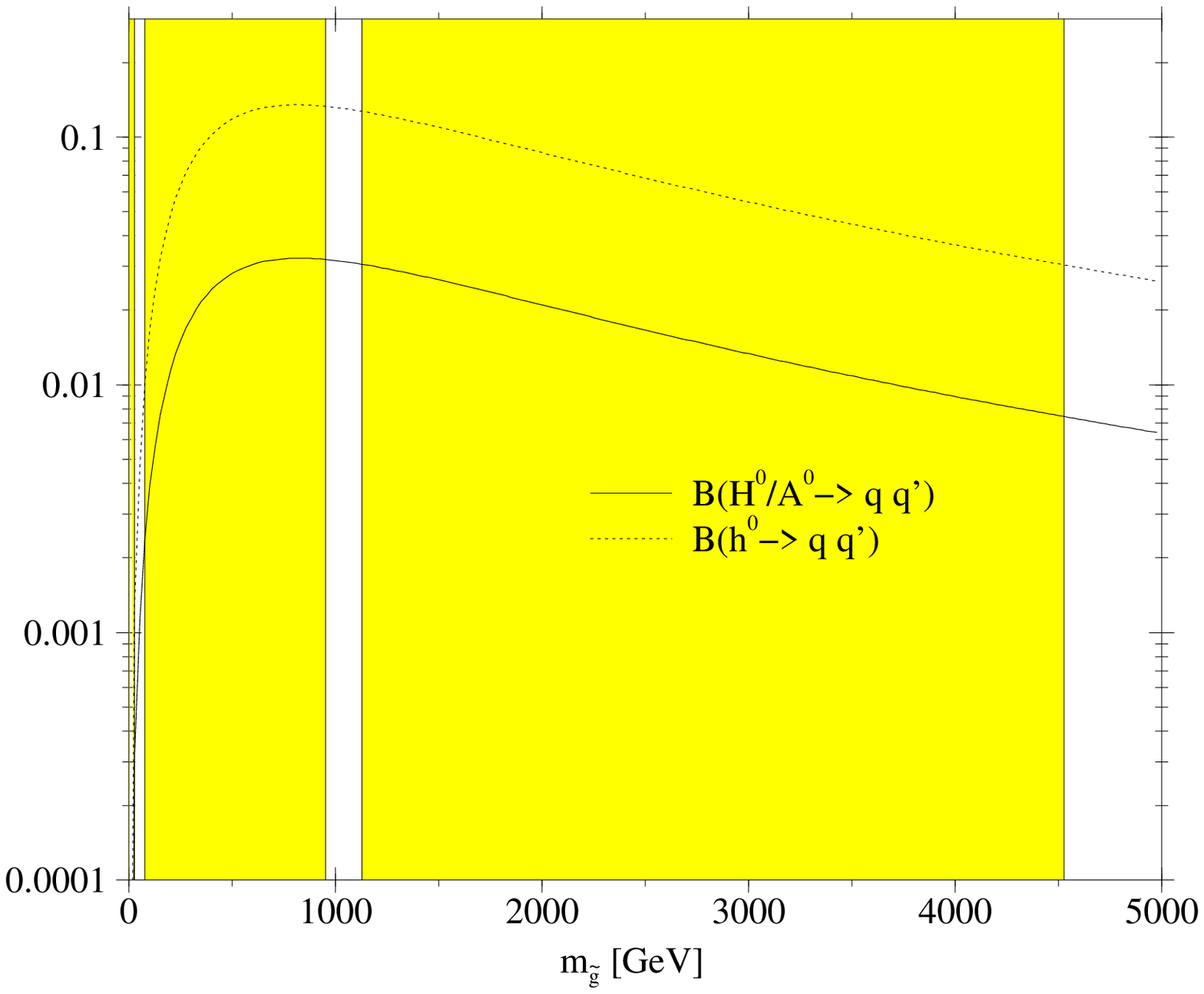}} &
\resizebox{!}{6cm}{\includegraphics{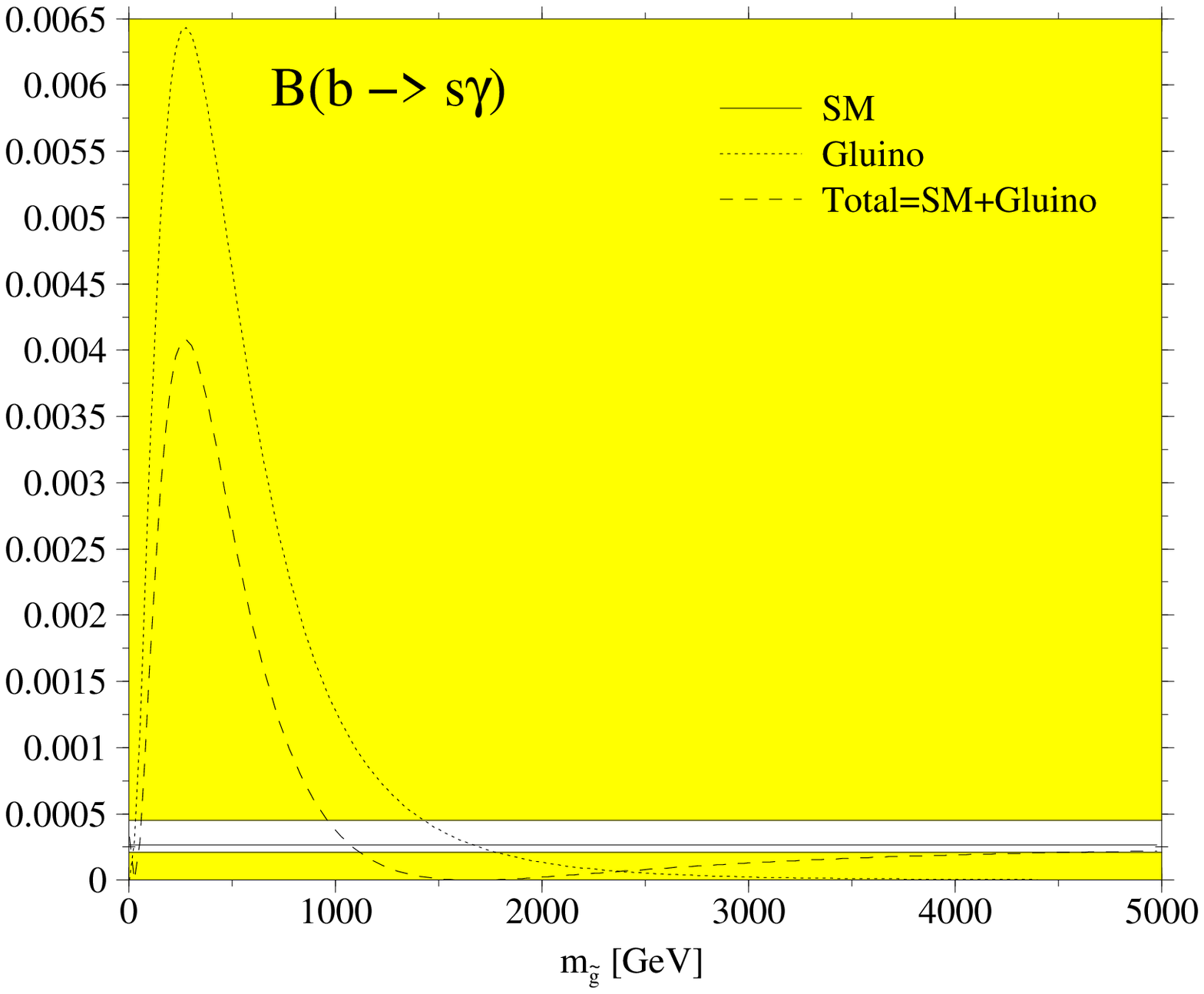}} \\
(a) & (b)
\end{tabular}
\caption{As in Fig.\,\ref{fig:mg}, but including the window region. {The remaining parameters are fixed as in the third
  column of Table~\protect{\ref{tab:maxim1}}.}}\label{fig:escletxes}
}%\end{figure}

{We further investigate the role of the scale of SUSY masses,
  and the fine-tuning behaviour in Figs.~\ref{fig:mg}
  and~\ref{fig:escletxes}. {In these figures we give up the equality
  $\mg=m_{\squark}$~(\ref{eq:scan-fixed}), the squark masses are fixed
  at the values stated in Tables~\ref{tab:maximnowindow}
  and~\ref{tab:maxim1} respectively.} Fig.~\ref{fig:mg} shows}
the
    values of $\Bhbs$ {for the three Higgs decays} and of $\Bbsg$
    as a function of the gluino mass for the
    parameters that maximize $\Bhzbs$ when the
    \textit{window} regions are excluded ({third column of
    Table~\ref{tab:maximnowindow}}). Here we see that, while the gluino
    contribution to $\Bbsg$ decouples \textit{fast} as a function of
    $\mg$, {its contribution} to $\Bhbs$ is fairly
    sustained. {Indeed},
    between $\mg=1\TeV$ and $\mg=5\TeV$ $\Bhzbs$ decreases
    only by a factor $\sim 1/4$, while the gluino contribution to $\Bbsg$
    becomes negligible
 at $\mg=5\TeV$ and we recover the SM
  prediction.
As a consequence, {the
    maximum rates $\Bhbs$ that we have found are robust}, in the sense that
    further theoretical refinements and experimental results that change
    the allowed range of $\Bbsg$ can easily {be compensated for by a slight
    increase} of the gluino mass ($\mg$), which would leave the
    prediction for $\Bhbs$ {essentially} unchanged.
    We note in Fig.\,\ref{fig:escletxes} the corresponding behaviour of
    $\Bhzbs$ and  $\Bbsg$  in the presence of fine-tuning, i.e. as in
    Table\,\ref{tab:maxim1}. {In contrast to} the previous
    case, here we observe the presence of  two tiny windows in the
    regions $\mg=25-75\GeV$ and $\mg=950-1125\GeV$. In the middle region
    $\mg=75-950\GeV$, $\Bbsg$ is one order of magnitude larger than the allowed
    experimental range, and in the region above $\mg=1125\GeV$ it only
    enters the allowed region for $\mg>4500\GeV$. In this region
    $\Bhzbs$ is still large, but at the price of having a gluino five
    times heavier than the rest of the SUSY spectrum. This is another
    manifestation of the large fine-tuning that governs this region of
    the parameter space.

\TABLE[pt]{%\begin{table}
\begin{tabular}{|c||c|c||c|c||c|c||}
\hline Particle & \multicolumn{2}{c||}{$H^0$} &
\multicolumn{2}{c||}{$h^0$} & \multicolumn{2}{c||}{$A^0$} \\
\hline & $\Gamma(\GeV)$ & \Bhbs & $\Gamma(\GeV)$ & \Bhbs &
$\Gamma(\GeV)$ & \Bhbs \\\hline
 \parbox{2cm}{\vspace{0.1cm} small-$\alpha_{eff}$ \textit{window}\vspace{0.1cm}}   &
11.0  & $ 3.3\times 10^{-2}$  & $ 1.6\times 10^{-3}$ & $
1.3\times 10^{-1}$  & 11.3  & $ 3.3\times 10^{-2}$\\\hline
 \parbox{2cm}{\vspace{0.1cm} tree-Higgs \textit{window} \vspace{0.1cm}} &
11.3 & $ 3.3\times 10^{-2}$ & $ 5.4\times 10^{-3}$ & $ 4.3\times
10^{-3}$ & 11.3 & $ 3.3\times 10^{-2}$\\\hline
\parbox{2cm}{\vspace{0.1cm} small-$\alpha_{eff}$ no-\textit{window}\vspace{0.1cm}}
& 11.2 & $ 9.1\times 10^{-4}$ & $ 1.4\times 10^{-3}$ & $
3.1\times 10^{-3}$ & 11.3 & $ 9.0\times 10^{-4}$\\\hline
 \parbox{2cm}{\vspace{0.1cm} tree-Higgs no-\textit{window} \vspace{0.1cm}}  &
11.3 & $ 9.1\times 10^{-4}$ & $ 5.4\times 10^{-3}$ & $ 1.3\times 10^{-4}$ & 11.3 & $ 9.0\times 10^{-4}$\\\hline
 $\tb=5$  &
0.11 & $ 2.0\times 10^{-3}$ & $ 6.0\times 10^{-3}$ & $ 1.7\times 10^{-4}$ & 0.11 & $ 2.1\times 10^{-3}$\\\hline
 \parbox{2cm}{\vspace{0.1cm}$\tb=5$ tree Higgs\vspace{0.1cm}}  &
0.12 & $ 1.9\times 10^{-3}$ & $ 4.4\times 10^{-3}$ & $ 2.6\times 10^{-4}$ & 0.11 & $ 2.1\times 10^{-3}$\\\hline
 \parbox{2cm}{\vspace{0.1cm}$\tb=5$ no-\textit{window}\vspace{0.1cm} }  &
0.15 & $ 3.8\times 10^{-4}$ & $ 9.7\times 10^{-3}$ & $ 1.1\times 10^{-4}$ & 0.11 & $ 5.1\times 10^{-4}$\\\hline
\end{tabular}
\caption{Maximum values of $\Bhbs$ and corresponding $\Gamma(h\to
X)$ for the different
  scenarios studied in this work.\label{tab:conclu}}
}%\end{table}

{Up to this point we have used the high $\tb$ value
{quoted in Eq.\,(\ref{eq:scan-fixed})}. But we have also looked at
the impact of varying $\tb$ on $\Bhbsmax$}. Since the latest LEP
data restricts $\tb\gsim 2.5$, we have used a moderate value of
$\tb=5$. Note that, at low $\tb$, the small $\alpha_{eff}$
scenario does not arise. {As a consequence similar results
are obtained} using either the tree-level or one-loop expressions
for the Higgs sector parameters. {We find that the three
branching ratios $\Bhbsmax$ at $\tb=5$ stay in the same order of
magnitude as in the scenarios with $\tb=50$ (default case) with
the tree-level Higgs sector and no-window (Cf.
Table\,\ref{tab:maximnowindow})}.

\section{Remarks and conclusions}
\label{sect:conclusions}

The main numbers of our analysis are put in a nutshell in
Table~\ref{tab:conclu}, {where} we show the results presented
previously, together with some other scenarios and the low $\tb$
case. The computed maximum values of $\Bhbs$ must
  not be taken as exact numbers {in practice}, but order of magnitude results. 
The implications that can be derived from
Table~\ref{tab:conclu} can be synthesized as follows:
\begin{enumerate}
\item The SUSY-QCD contributions can enhance  the maximum
  expectation for the FCNC decay rates $\Bhbs$ {enormously}. {This
  is seen by comparing the results of Table~\ref{tab:conclu} with the maximum
  value of $B(H^{SM}\rightarrow b\bar{s})$
  considered in Eq.\,(\ref{estimateBRSM})}. The optimized MSSM
  branching ratios are at the very least 3 orders of magnitude
  bigger than the SM result.

\item If no special circumstances apply, that is, {if} no fine-tuning occurs
  between the parameters contributing to $\Bbsg$ in the MSSM, and {if} $\Gamma(h^0\to b\bar{b})$ is not
  suppressed, the maximum
  rates are $\Bhzbsmax\simeq 1.3\times 10^{-4}$, $\BHAbsmax\simeq 9\times 10^{-4}$. This corresponds to the
  {\textit{tree-Higgs}/\textit{no-window}} scenario in
  Table~\ref{tab:conclu}.
  \label{sc:noloopnofine}
\item If, however, $\Gamma(h^0\to b\bar{b})$ is suppressed by the radiative corrections
to the CP-even mixing angle $\alpha$, then $\Bhzbs$ can be an
order of magnitude larger: $\Bhzbsmax\sim 3\times 10^{-3}$. This
corresponds to the
  small $\alpha_{eff}$ scenario, and is indicated by small-$\alpha_{eff}$/ no-\textit{window} in
  {Table}~\ref{tab:conclu}. The FCNC branching ratio that we find for $h^0$ in this case should
  be considered as the largest possible one within the conditions of naturalness (no
  fine-tuning).
  \label{sc:loopnofine}
\item On the other hand, if fine-tuning between the gluino and the SM contributions to $\Bbsg$
  is allowed, but the small-$\alpha_{eff}$ scenario is not realized,
  then $\Bhzbsmax$ grows one order of
  magnitude up to $\Bhzbsmax\sim 4\times 10^{-3}$, whereas $\BHAbsmax\sim
  3\times 10^{-2}$. {This corresponds to the case labelled
  \textit{tree-Higgs/window} in Table~\ref{tab:conclu}}.
  \label{sc:noloopfine}
\item When both special conditions take place simultaneously, viz. fine-tuning in
$\Bbsg$ (triggered by a very special choice of the $\delta_{23}$
parameter in a narrow window range) and small $\alpha_{eff}$
scenario (independent of assumptions on $\delta_{23}$), we reach
an over-optimistic situation where $\Bhzbsmax$ could reach the
$\sim 10\%$ level. This is the case referred to as
small-$\alpha_{eff}$/ \textit{window} in Table~\ref{tab:conclu}.
  \label{sc:loopfine}
  \item If $\tb$ is low/moderate, then $\Bhbsmax$ lie in the lower
  range $\sim 10^{-4}$, which can grow an order or magnitude for
  $\BHAbsmax$ in fine-tuned scenarios (last three rows in
  Table~\ref{tab:conclu}).
\end{enumerate}

Although the large FCNC rates mentioned in points
\ref{sc:noloopfine} and \ref{sc:loopfine} above seem to offer a
rather tempting perspective, we will not elaborate on them any
further since in our opinion the fine-tuning requirement inherent
in them is too contrived. 
On the other hand, points~\ref{sc:noloopnofine}
and~\ref{sc:loopnofine} offer a moderate, but certainly much more
realistic scenario, which in no way frustrates our hopes to
potentially detect the FCNC Higgs boson decays (\ref{hFCNC}). 
Indeed, in the case described in point \ref{sc:noloopnofine}, $\Bhbs$
can be at most of order $10^{-4}$.
But this is still a fairly
respectable FCNC branching ratio (comparable to that of
$b\rightarrow s\gamma$) and it may lead to a large number of
events at a high luminosity collider\,\cite{Future}\footnote{{See
e.g. Ref.\cite{Bejar:2003em} for a detailed analysis of the
number of Higgs boson FCNC events produced at the LHC in a
different situation corresponding to the general 2HDM.}}.
Moreover, if $\Gamma(h\to X)$ becomes suppressed (e.g. by
realizing the small $\alpha_{eff}$ scenario, point~\ref{sc:loopnofine})
then $\Bhzbs$ can be 
enhanced {by} an additional order of magnitude. 

{Our analysis correlates the values of $\Bhbs$ with that of
  $\Bbsg$, taking into account only the SUSY-QCD contributions due to
  flavour mixing parameters among the left-chiral squarks. The presence
  of several other competing contributions to $\Bbsg$ alters the borders
  of the allowed parameter space:}
\begin{itemize}
\item {For
  the fine-tuned scenarios, the presence and
  position of the allowed \textit{window} regions in the parameter space
  depends significantly on all the contributions, and therefore also
  does the maximum value of $\Bhbs$. Outside the \textit{window} regions,
  the computed value of $\Bbsg$ can only be made consistent with the
  experimental range, by means of a large splitting between the squark
  and gluino  masses.}
\item {For the non-fine-tuned scenarios, the
  inclusion of further contributions to $\Bbsg$ also alters the allowed
  parameter space, but the condition of non-fine-tuning ensures
  precisely that the change in the allowed range of $\delta_{23}$ is
  smooth, and the corresponding change in $\Bhbsmax$ is not dramatic. }
\end{itemize}

Of course, the question immediately arises on what will happen if
the data from present $B$-meson factories further constrains the
$\delta_{23}$ parameter. In that case, we should take into
account the (charged-current induced) SUSY-EW contributions to
$\Bhbs$, which will be presented in Ref.~\cite{Future}
(see also \cite{Madrid2}). However, we can advance that
the SUSY-EW effects on $\Bhbsmax$ that we find are in
the ballpark of $\Bhzbsmax\sim 3 \times 10^{-5}$ and
$\BHAbsmax\sim 1\times10^{-5}$ for a non-fine-tuned scenario,
{while} $\Bhzbsmax \sim 2\times 10^{-4}$ and $\BHAbsmax\sim
8\times 10^{-5}$ for a fine-tuned scenario. {From the analysis of
Ref.\,\cite{Bejar:2003em} we expect that even with these
impoverished MSSM rates the number of FCNC events of that sort
should be non-negligible at the LHC.}

We have already mentioned that our results disagree some orders
of magnitude with {recent} estimates presented in the
literature~\cite{Madrid}. {In fact, in our analysis we
cannot accommodate a branching ratio}
{at the level of $\Bhbs\sim (20-30)\%$} for any of the
decays (\ref{hFCNC}), as claimed by these authors. We find such
values incompatible with a {rigorous} MSSM analysis of
these decays correlated with the branching ratio of $\bsg$.
{Even though we have detected the existence of corners of
the MSSM parameter space where a Higgs boson FCNC branching ratio
can barely reach the $10\%$ level (cf. the narrow windows in
Fig.\,\ref{fig:d23bsg} and Fig.\,\ref{fig:escletxes}), we insist
once more that they should be considered rather unlikely as they
are associated to fine tuning of the parameters. Moreover, in
contrast to these authors, we find that it is the lightest
CP-even state, $h^0$, the one that could have the largest FCNC
branching ratio. Thus, as already advanced in
Ref.\,\cite{Bejar:2003em}, we believe that the authors of
Ref.\,\cite{Madrid} have overestimated by a significant amount
the value of $\Bhbsmax$ for the three Higgs bosons of the MSSM}.

{To conclude, we have presented a first realistic estimate of
the branching ratios of the Higgs boson FCNC decays (\ref{hFCNC})
within the MSSM, assuming that the SUSY-QCD corrections can be as
large as permitted by the experimental constraints on $\Bbsg$.
{We have carried out a systematic and self-consistent
maximization of the branching ratios (\ref{eq:hbs-def}) taking
into account this crucial experimental constraint. At the end of
the day the results that we obtain, especially for the lightest
CP-even Higgs boson of the MSSM, are fairly large: $\Bhzbsmax\sim
10^{-4}-10^{-3}$}. These MSSM rates turn out to be {between three
to four orders of magnitude} larger than the maximum SM rate
(\ref{estimateBRSM}), but not five or six orders as naive
expectations indicated. Whether this branching ratio is
measurable at the LHC~\cite{\LHC} or at a high energy $e^+e^-$
Linear Collider~\cite{\TESLA} can only be established by means of
specific experimental analyses. However, on the basis of related
studies in the general 2HDM\,\cite{Bejar:2003em} and from ongoing
work in the MSSM\,\cite{Future}, we can foresee that an important
number of FCNC events (\ref{hFCNC}) can be potentially collected
at the LHC. {They could play a complementary, if not decisive,
role in the identification of low-energy Supersymmetry}. In this
paper we have dealt only with the maximum rates induced by the
SUSY-QCD sector of the model. A more detailed analysis --
{including the SUSY-EW sector and the computation of the
aforementioned production rates} -- will be presented in a
forthcoming publication~\cite{Future}}.

%\section*{Acknowledgments}
\acknowledgments
J.G. thanks J{\"o}rg Urban for useful discussions on $\bsg$. 
This collaboration is part of the network ``Physics at
Colliders'' of the European Union under contract
HPRN-CT-2000-00149. The work of S.B. has been supported in part
by CICYT under project No. FPA2002-00648, and that of F.D. and J.S.  by
MECYT and FEDER under project FPA2001-3598.
 F.D. has also been
supported by the fellowship 2003FI 00547 of the Generalitat de Catalunya.
\providecommand{\href}[2]{#2}

\end{document}